\documentclass[11pt]{article}
\usepackage[final]{acl}
\usepackage{times}
\usepackage{latexsym}
\usepackage[T1]{fontenc}
\usepackage[utf8]{inputenc}
\usepackage{microtype}
\usepackage{inconsolata}
\usepackage{graphicx}

\usepackage{pdfpages}
\usepackage{graphicx}
\usepackage{longtable}
\usepackage{tabularx}
\usepackage{booktabs}

\usepackage{amsmath}
\usepackage{amssymb}
\usepackage{algorithm}
\usepackage{algorithmic}

\usepackage{tcolorbox}

\usepackage{mdframed}
\title{From Isolated Scoring to Collaborative Ranking: \\A Comparison-Native Framework for LLM-Based Paper Evaluation}

\author{
 \textbf{Pujun Zheng\textsuperscript{1}},
 \textbf{Jiacheng Yao\textsuperscript{1}},
 \textbf{Jinquan Zheng\textsuperscript{1}},
 \textbf{Chenyang Gu\textsuperscript{1}},
 \textbf{Guoxiu He\textsuperscript{1}}\thanks{Corresponding author.}, \\
 \textbf{Jiawei Liu\textsuperscript{2}},
 \textbf{Yong Huang\textsuperscript{2}},
 \textbf{Tianrui Guo\textsuperscript{3}},
 \textbf{Wei Lu\textsuperscript{2}}
\\
 \textsuperscript{1}School of Economics and Management, East China Normal University,\\
 \textsuperscript{2}School of Information Management, Wuhan University,\\
 \textsuperscript{3}China Academic Degrees \& Graduate Education Development Center
\\
\texttt{\{pjzheng, jcyao, jqzheng, cygu\}@stu.ecnu.edu.cn, gxhe@fem.ecnu.edu.cn} \\
}

\begin{document}

\maketitle

\begin{abstract}
	Large language models (LLMs) are currently applied to scientific paper evaluation by assigning an absolute score to each paper independently. However, since score scales vary across conferences, time periods, and evaluation criteria, models trained on absolute scores are prone to fitting narrow, context-specific rules rather than developing robust scholarly judgment. To overcome this limitation, we propose shifting paper evaluation from isolated scoring to collaborative ranking. In particular, we design a \textbf{C}omparison-\textbf{N}ative framework for \textbf{P}aper \textbf{E}valuation (\textbf{CNPE}), integrating comparison into both data construction and model learning. We first propose a graph-based similarity ranking algorithm to facilitate the sampling of more informative and discriminative paper pairs from a collection. We then enhance relative quality judgment through supervised fine-tuning and reinforcement learning with comparison-based rewards. At inference, the model performs pairwise comparisons over sampled paper pairs and aggregates these preference signals into a global relative quality ranking. Experimental results demonstrate that our framework achieves an average relative improvement of \textbf{21.8\%} over the strong baseline DeepReview-14B, while exhibiting robust generalization to five previously unseen datasets. Our code is available at \href{https://github.com/ECNU-Text-Computing/ComparisonReview}{github}.
\end{abstract}

\section{Introduction}

\begin{figure}[htbp]
	\setlength{\belowcaptionskip}{-15pt} 
	\setlength{\floatsep}{10pt} 
	\setlength{\textfloatsep}{10pt}
	\centering
	\includegraphics[width=\columnwidth]{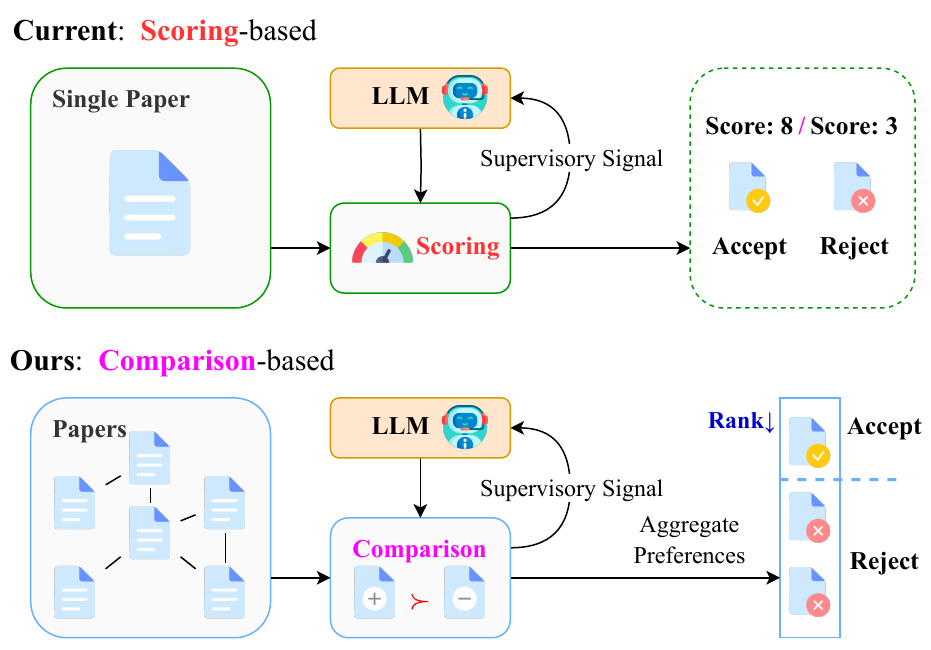}
	\caption{Comparison of current methods (top) and our approach (bottom) across data (left), training (middle), and inference (right).}
	\label{fig:methodology}
\end{figure}

Paper evaluation plays a central role in advancing scientific progress \cite{margolis1967citation}. Peer review has long served as the primary mechanism for ensuring publication quality and rigor \citep{alberts2008reviewing}. However, the rapid growth of submissions across disciplines has placed increasing pressure on the peer review system \cite{he2023research}, exacerbating issues such as inconsistent evaluations and inherent biases \cite{xue2023re,lee2013bias,tomkins2017reviewer}. In recent years, large language models (LLMs) \cite{naveed2023comprehensive} have been increasingly applied to paper evaluation, enabling more consistent and scalable review processes \citep{du2024llms, zhou2024llm, zhuang2025large, thakkar2025can}. However, their growing adoption and societal impact do not eliminate the significant limitations of LLMs \citep{latona2024ai, liang2024monitoring, zhu2025your}.

Currently, LLM-based paper evaluation methods primarily assign scores to individual papers (as illustrated at the top of Figure \ref{fig:methodology}). As closed knowledge systems, LLMs have inherent limitations in performing the complex reasoning and judgment required for cutting-edge scientific innovation \citep{liu2023reviewergpt, purkayastha2025lazyreview}, making them prone to hallucinations and retrieval biases \citep{li2025can}. Multi-agent frameworks that simulate human peer review workflows cannot fundamentally resolve these issues and may inherit biases present in manual review, such as favoring predictable or mediocre results \citep{zhang2025from}. Even when models are trained to align with human-assigned absolute scores \citep{lu2024ai, yu2024automated, weng2025cycleresearcher, zhu2025deepreview}, performance is constrained by dataset-specific factors: score scales vary across conferences, time periods, and evaluation criteria, causing models to learn context-specific rules rather than generalizable scholarly judgment. Although several recent studies use pairwise or listwise comparisons \citep{zhang2025from, zhao2025words}, they either do not train model parameters for comparison tasks or still output absolute scores, limiting both performance and generalization \cite{hopner2025automatic}. Consequently, existing methods are impractical and fail to produce reliable, consistent rankings among papers, such as a batch of conference submissions.

To this end, we propose a comparison-native framework that uses collaborative data to train the LLM to evaluate and rank papers (as illustrated at the bottom of Figure \ref{fig:methodology}). Unlike methods that assign absolute scores to individual papers, our framework reframes the complex task of quantitative reasoning as a comparison problem, better leveraging the reasoning strengths of LLMs \cite{wei2022chain}. By learning more reliable preference signals, the model can more clearly capture differences across novelty, significance, and clarity. Furthermore, the comparative logic acquired through this approach is more transferable than that of single-paper scoring, enabling generalization to new paper collections without adapting to specific scoring scales.

Specifically, we design two key modules for effective comparison: pair sampling and quality judgment. At the data level, pair sampling selects the most informative paper pairs from all possible combinations to ensure diverse contextual coverage, enhancing comparison accuracy and model generalization. We facilitate this using a graph-based similarity ranking algorithm that prioritizes papers with overlapping research areas. 
At the model level, quality judgment assesses candidate papers along dimensions such as novelty, significance, and clarity, reliably identifying superior papers and handling fine-grained distinctions. To strengthen judgment capabilities, we construct a dataset of paper pairs annotated with quality preferences and optimize the model via supervised fine-tuning combined with reinforcement learning using verifiable comparison-based rewards. At inference, the model performs pairwise comparisons over sampled paper pairs and aggregates the resulting preference signals into an interpretable global ranking of relative paper quality.

Our framework achieves leading performance on the ICLR-2025 dataset for both paper quality ranking and acceptance prediction across all evaluation metrics. With only 7B parameters, it achieves an average relative improvement of 21.8\% over the strong baseline DeepReview-14B \citep{zhu2025deepreview}. Ablation studies validate the effectiveness of our training strategy based on supervised fine-tuning and reinforcement learning, and reveal that jointly applying similarity-based and random sampling enhances performance. The model exhibits strong generalization to five unseen submission datasets from top-tier conferences in 2025, including ICML, NeurIPS, ACL, EMNLP, and NAACL. Our main contributions are as follows:

$\bullet$ We propose an LLM-based comparison-native framework that implements paper ranking through data construction and model training.

$\bullet$ We propose a graph-based semantic similarity recognition method for paper pair sampling and a comparison-based reward to enhance quality judgment through reinforcement learning.

$\bullet$ Our framework achieves leading performance in ranking and acceptance prediction, with robust generalization to previously unseen datasets.

\section{Related Work}
\label{sec:relate}

\paragraph{Agent‑based Assessment Systems}

These systems typically combine intelligent agents with general-purpose LLMs to generate review outcomes. \citet{jin2024agentreview} employed an LLM-powered peer-review simulation framework to replicate the review process and identify potential influencing factors. \citet{lu2024ai} designed automated review systems to simulate peer evaluation in paper scoring tasks. Other notable efforts include customizing feedback through novel alignment mechanisms with iterative optimization \cite{garg2025revieweval}, implementing tree-structured workflows to enhance performance \cite{chang2025treereview}, and creating multi-agent frameworks that merge prompt engineering, collaboration, shared memory, and multimodal perception to deliver high-quality reviews \cite{lu2025agent}. 
However, these methods generally rely on LLMs that have not been specifically trained for academic review, which limits their reliability, while the reliance on general-purpose models increases the risk of data leakage.

\paragraph{Training-based Review Models}

These approaches transform open-source LLMs into expert reviewers through domain-specific fine-tuning or reinforcement learning. \citet{yu2024automated} developed a domain-specific review system. \citet{tan2024peer} introduced a multi-turn dialogue mechanism to capture the dynamic nature of reviews. \citet{tyser2024ai} fine-tuned LLMs to predict human-preference-aligned evaluations. \citet{weng2025cycleresearcher} combined manuscript generation with iterative review to facilitate scientific discovery, and \citet{zhu2025deepreview} designed a multi-stage workflow that integrated structured analysis, literature retrieval, and evidence-based reasoning to achieve optimal performance. \citet{zeng2025reviewrl} introduced a reinforcement learning framework for review generation. Other fine-grained improvements include structured reasoning \cite{dycke2025stricta}, avoiding lazy thinking \cite{purkayastha2025lazyreview}, and identifying research limitations \cite{xu2025llmsidentify}. 
Despite performance gains from training, most current methods rely on absolute scores for individual papers as supervisory signals, which tend to overfit superficial patterns and limit generalization.

\paragraph{Comparison-Based Evaluation Methods}

These methods advance the review process through pairwise or listwise strategies. This concept originated in early work~\cite{cao2007learning} and, in the era of LLMs, has been extended to domains such as essay scoring~\cite{shibata2025lces} and LLM evaluation~\cite{ning2025pico}. For paper review, \citet{zhang2025from} performed pairwise comparisons between papers and aggregated the resulting preferences to reconstruct a robust global ranking; however, the LLM is treated as a fixed comparator without task-specific training for comparison, leaving its judgments constrained by general-purpose priors and prompt design. \citet{zhao2025words} predicted normalized impact scores from titles and abstracts; although their training relied on listwise comparisons, inference still returned to predicting an isolated absolute score for each paper via a black-box process. \citet{hopner2025automatic} reported that pairwise ranking prediction encounters substantial difficulties in estimating review scores, and that various attempted improvements have yielded little success. Consequently, although comparative reasoning is partially incorporated in existing methods, systematically strengthening comparison-native modeling from the perspectives of data construction, model learning, and inference remains largely unexplored yet highly promising. 

\begin{figure*}[htbp]
	\setlength{\belowcaptionskip}{-5pt} 
	\setlength{\floatsep}{10pt} 
	\setlength{\textfloatsep}{10pt}
	\centering
	\includegraphics[width=\textwidth]{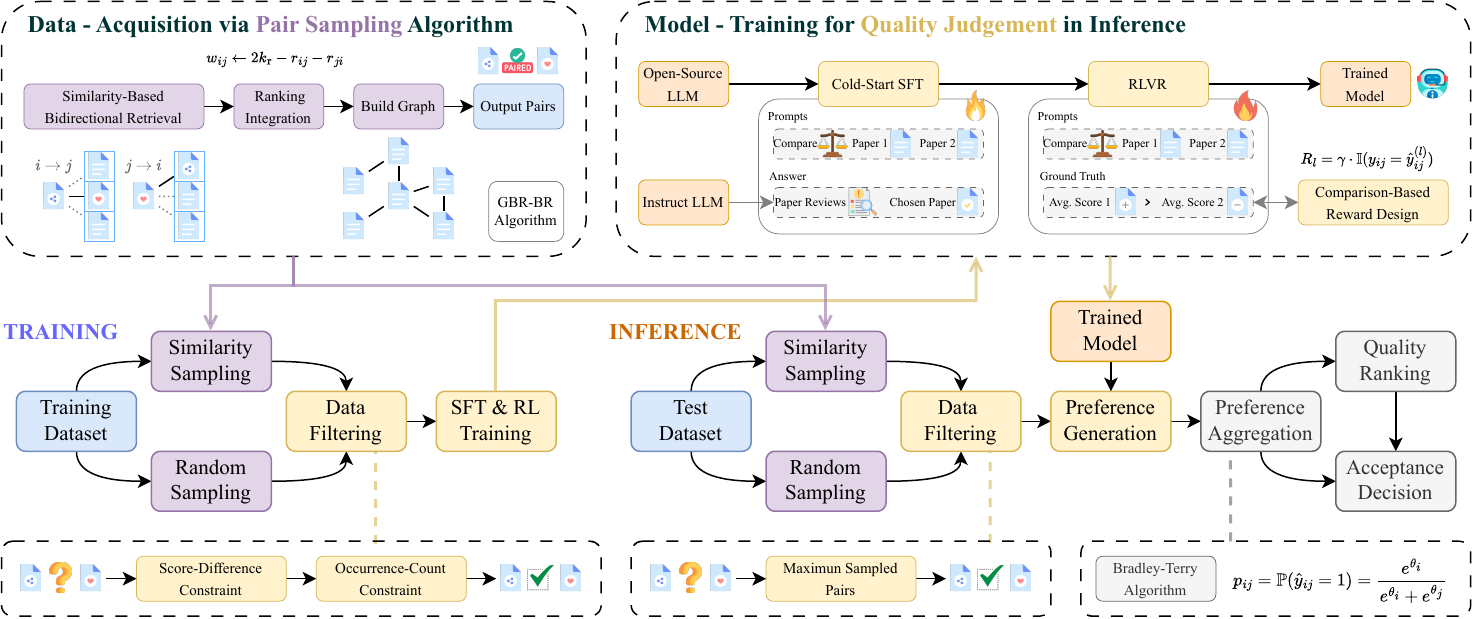}
	\caption{Overview of the framework. During training (bottom left), pair sampling (top left) provides sample pairs for learning quality judgment (top right). At inference (bottom right), pair sampling supplies pairs for quality judgment by the trained model, followed by ranking and decision based on preference.}
	\label{fig:graph}
\end{figure*}

\section{Methodology}
\label{sec:method}

\subsection{Overview} 

Figure \ref{fig:graph} shows the proposed comparison‑native LLM framework for paper evaluation. The framework applies a consistent pair‑sampling strategy for both training and inference, selecting in‑domain and cross‑domain paper pairs. During training, pairs are filtered by score difference and occurrence constraints to retain informative comparisons, and the LLM is optimized through comparison‑based reinforcement learning. During inference, pair selection is limited by an upper‑bound constraint, and the model's pairwise preferences are aggregated via the Bradley-Terry model to yield an overall ranking.

\subsection{Pair Sampling}
\label{sec:sampling}

\begin{algorithm}[ht]
	
	\caption{Graph-based Ranking with Bidirectional Retrieval (GBR-BR)}
	\label{alg:gbr_br}
	\begin{algorithmic}[1]
		\small
		\REQUIRE Paper set $\mathcal{P} = \{p_1, p_2, \dots, p_n\}$, embedding model $\mathrm{Embed}$, reranking model $\mathrm{Rerank}$, top-$k$ parameters $k_{\mathrm{e}}, k_{\mathrm{r}}$.
		\ENSURE Sorted list of paper pairs $S$
		
		\STATE Initialize $w_{ij} \gets 0$, $r_{ij} \gets \infty$ for all $i,j  \in \{1, \dots, n\}$
		\FOR{each $p_i \in \mathcal{P}$}
		\STATE $C_i \gets$ Top-$k_{\mathrm{e}}$ candidates from $\mathrm{Embed}(p_i)$
		\FOR{each $p_j \in C_i$}
		\STATE $r_{ij} \gets$ rank of $p_j$ in $\mathrm{Rerank}(p_i, C_i, k_{\mathrm{r}})$
		\ENDFOR
		\ENDFOR
		\FOR{$i \in \{1, \dots, n\}$}
		\FOR{$j \in \{i+1, \dots, n\}$}
		\IF{$r_{ij} < k_{\mathrm{r}}$ \OR $r_{ji} < k_{\mathrm{r}}$}
		\STATE $w_{ij} \gets 2k_{\mathrm{r}} - r_{ij} - r_{ji}$
		\ENDIF
		\ENDFOR
		\ENDFOR
		\STATE Construct graph $\mathcal{G}  =  (\mathcal{P}, \{(i,j) : w_{ij} > 0\}, w)$
		\STATE Check connectivity of $\mathcal{G}  \rightarrow \texttt{is\_connected}$
		\IF{$\neg$ $\texttt{is\_connected}$}
		\STATE Increase $k_{\mathrm{e}}$ and $k_{\mathrm{r}}$, \textbf{goto} step 1
		\ENDIF
		\STATE $S \gets \text{sort}(\{(i,j)\}, -w_{ij})$
		\RETURN $S$
	\end{algorithmic}
\end{algorithm}

The comparison-native framework relies on training and inference data organized as comparison pairs. Randomly sampling paper pairs from a corpus makes it difficult to obtain both in-domain and cross-domain comparisons. We therefore propose a pair sampling strategy applied before training set selection and evaluation pair construction.
\paragraph{Domain-aware Pair Ranking}
To ensure in-domain comparability, sampled paper pairs are required to originate from closely related research domains. However, domain boundaries have become increasingly fluid, with frequent cross-domain overlap, rendering rigid and static domain divisions difficult to operationalize and prone to bias by ignoring domain intersectionality. This limits the effectiveness of existing domain classification schemes in improving ranking accuracy \citep{hopner2025automatic}. 
Unlike treating domains as isolated clusters, we explicitly model pairwise comparability between papers. We construct a weighted sparse graph to encode semantic similarity, with edges representing candidate paper pairs. To avoid isolated papers and favor comparisons within similar domains, the graph must satisfy two conditions: (1) all nodes are connected; and (2) semantically similar paper pairs are assigned higher edge weights.
To assign weights to candidate edges in the paper graph, we propose Graph-based Ranking with Bidirectional Retrieval (GBR-BR), as shown in Algorithm~\ref{alg:gbr_br}.
For each paper, an embedding model retrieves a set of semantically relevant candidates, which are then reranked to form an ordered list reflecting their importance. Since a paper may appear in both its own list and another paper's list, inconsistencies can arise between the two rankings. We address this by bidirectionally integrating the rankings from two lists obtained via retrieval, which mitigates the asymmetry of unidirectional retrieval and ensures consistent edge weights. Edge weights are then assigned based on the integrated ranking results, with higher weights for more semantically similar pairs, whereas lower-ranked or irrelevant pairs are discarded. The remaining edges form a weighted sparse graph, which is checked for connectivity; if isolated nodes exist, the number of candidates or ranking thresholds is increased and the process is repeated to ensure all papers are connected. Finally, the paper pairs are sorted by edge weights to produce a high-quality, informative set of comparison pairs for both training and inference.

\paragraph{Sampling Strategy}
To satisfy the requirements of both in-domain and cross-domain comparison, we employ two complementary sampling strategies. (1) Similarity-based sampling applies proximity matching using the GBR-BR algorithm to construct paper pairs from semantically similar domains. These pairs emphasize fine-grained quality comparison and constitute the primary source of supervision for learning paper quality judgment. 
(2) Random-based sampling generates paper pairs via random matching without enforcing semantic or topical relevance, using a greedy procedure to avoid duplication. This strategy provides cross-domain comparisons that enhance the model's ability to generalize across diverse research areas. 

Training and inference sets are created by further filtering the pairs generated through these two sampling strategies. The specific filtering procedures are described in the relevant sections below.

\subsection{Training}
\label{sec:training}

Training plays a central role in the comparison-native framework, as it equips the model with the ability to make precise and reliable pairwise comparisons between papers.

\paragraph{Training Pairs Filtering}
The construction of paper pairs for training is subject to two constraints. (1) Score-difference constraint. Let $s_i$ be the average ground-truth score of paper $i$. This constraint specifies that a pair is considered valid only when the score difference exceeds a predefined threshold. This reduces the noise introduced by pairs with similar scores, thereby enabling the model to learn stronger and more informative signals. (2) Occurrence-count constraint. Let $\mathrm{count}_i$ denote the number of times paper $i$ appears in the sampled pairs. By limiting $\mathrm{count}_i$, the model is encouraged to learn from a more diverse set of supervision signals rather than repeatedly seeing the same samples, which in turn improves its generalization ability. The complete filtering rule for training paper pairs is formulated as:
\begin{equation}
	\Phi_{\text{train}} := 
	\begin{cases}
		\forall i \neq j,\; |s_i - s_j| \geq d_{\min} \\
		\forall i,\; \mathrm{count}_i \leq c_{\max}
	\end{cases}
\end{equation}

\paragraph{Cold-Start SFT and RLVR}

We first construct reasoning chains for distinguishing paper quality using an Instruct LLM with the titles and abstracts of papers, performing cold-start via supervised fine-tuning. 
Subsequently, we employ reinforcement learning to enhance the LLM's reasoning ability for comparison tasks.
Specifically, we adopt an improved GRPO \cite{yu2025dapo,liu2025understanding} for reinforcement learning. The optimization objective is:

\begin{equation}
	\begin{aligned}
		\mathcal{J}  & (\theta)= \mathbb{E}_{(q,a)\sim\mathcal{D},\,\{o_{l}\}_{l=1}^{G}\sim\pi_{\theta_\mathrm{old}}(\cdot|q)} \\
		& \Bigg[ 
		\frac{1}{G} \sum_{l=1}^{G} \sum_{t=1}^{|o_{l}|} 
		\min\Big(  r_{l,t}(\theta)\hat{A}_{l,t}, \\
		& \mathrm{clip}\big(r_{l,t}(\theta),\,1-\varepsilon_{\mathrm{low}},\,1+\varepsilon_{\mathrm{high}}\big)\hat{A}_{l,t} \Big)
		\Bigg]
	\end{aligned}
\end{equation}

Where:

\begin{equation}
	r_{l,t}(\theta)=\frac{\pi_{\theta}(o_{l,t}\mid q,o_{l,<t})}{\pi_{\theta_{\mathrm{old}}}(o_{l,t}\mid q,o_{l,<t})}
\end{equation}

And:

\begin{equation}
	\hat{A}_{l,t}={R_{l}-\mathrm{mean}(\{R_{l}\}_{l=1}^{G})}
\end{equation}

We use a comparison-based reward mechanism that issues rewards only when the model produces correct comparison outcomes, in contrast to existing RL approaches that derive rewards directly from numeric scores \cite{zeng2025reviewrl}. To overcome the high cost and impracticality of collecting human feedback, we employ a verifiable, rule-driven reward strategy that leverages the actual mean review scores from original ICLR submissions. The model's comparison accuracy is assessed by evaluating whether the mean score of one paper exceeds that of another.

Formally, given the human-annotated mean scores $s_i$ and $s_j$ from the authentic reviews, the ground-truth comparison label $y_{ij}$ is defined as:
\begin{equation}
	y_{ij} = \mathbb{I}\big(s_i > s_j \big)
\end{equation}
The prediction generated by the LLM in the $l$-th rollout is denoted as $\hat{y}_{ij}^{(l)}$. The notation $p_i \succ p_j$ indicates that the model considers $p_i$ to have higher quality than $p_j$:
\begin{equation}
	\hat y_{ij}^{(l)} = \mathbb{I}\big( f_{\text{LLM}}^{(l)}(p_i, p_j) = p_i \succ p_j \big)
\end{equation}
The reward signal is obtained by comparing the predicted preference with the ground-truth label. Here, $\gamma$ is a positive scalar controlling the reward magnitude:
\begin{equation}
	R_l = \gamma \cdot \mathbb{I}(y_{ij} = \hat{y}_{ij}^{(l)})
\end{equation}

\subsection{Inference}
\label{sec:inference}

During inference, given $n$ papers, there are theoretically a quadratic number of possible paper pairs.

\paragraph{Inference Pairs Filtering}
After applying the pair sampling strategy, the filtering objective for inference is to ensure that all papers are covered. We control the number of pairs in this set by selecting a fraction $\alpha$ relative to the theoretical total. 

\paragraph{Preference Generation and Aggregation}

For each paper pair $(i, j)$, the trained LLM generates a preference label $\hat y_{ij}$. To aggregate these pairwise preferences into an overall ranking, we associate each paper with a latent quality score $\theta_i \in \mathbb{R}$. The probability that one paper is preferred over another is modeled using the Bradley-Terry model, which defines the probability of paper $i$ being preferred over paper $j$ as: 
\begin{equation}
	p_{ij} = \mathbb{P}(\hat y_{ij} = 1) = \frac{e^{\theta_i}}{e^{\theta_i} + e^{\theta_j}}
\end{equation}
Given all observed pairwise preference labels, the total log-likelihood over all ordered pairs is:
\begin{equation}
	\mathcal{L}_{\theta} = \sum_{i \ne j} \big[ \hat y_{ij} \log p_{ij} + (1 - \hat y_{ij}) \log (1 - p_{ij}) \big] 
\end{equation}
Maximizing this log-likelihood yields estimates of the latent quality scores for all papers, from which we derive a descending ranking. Papers whose ranks are above a predefined threshold are accepted, while the others are rejected.

\begin{table*}[t]
	\setlength{\belowcaptionskip}{-10pt} 
	\setlength{\floatsep}{10pt} 
	\setlength{\textfloatsep}{10pt}
	\centering
	\scriptsize
	\setlength{\tabcolsep}{3pt}
	\begin{tabularx}{\textwidth}{@{}
			>{\raggedright\arraybackslash}p{0.14\linewidth}
			*{9}{>{\centering\arraybackslash}X}@{}} 
		\toprule
		& \multicolumn{4}{c}{\textbf{Decision}} & \multicolumn{4}{c}{\textbf{Ranking}} &  \\
		\cmidrule(lr){2-5} \cmidrule(lr){6-9}
		\textbf{Method} & \textbf{Accuracy} & \textbf{F1} & \textbf{AUC} & \textbf{Cohen \(\kappa\)} & \textbf{Spearman \(\rho\)} & \textbf{Pair. Acc.} & \textbf{MAP@20} & \textbf{NDCG@20} & \textbf{Avg. Perf.} \\
		\midrule
		\multicolumn{10}{@{}l}{\textbf{pointwise - agents}} \\
		AIScientist(GPT) & \underline{0.6972} & 0.5301 & 0.6449 & 0.1246 & 0.3106 & 0.6045 & \underline{0.4637} & \underline{0.7863} & 0.7879 \\
		AIScientist(Gemini) & 0.5615 & 0.5397 & 0.5700 & 0.1038 & 0.1659 & 0.5562 & 0.0966 & 0.6464 & 0.6061 \\
		AIScientist(GLM) & 0.3801 & 0.3516 & 0.5912 & 0.0479 & 0.3029 & 0.5865 & 0.2200 & 0.7091 & 0.6022 \\
		AgentReview(GPT) & 0.5079 & 0.5003 & 0.5506 & 0.0628 & 0.0503 & 0.5187 & 0.2320 & 0.6674 & 0.5560 \\
		AgentReview(Gemini) & 0.5379 & 0.5258 & 0.5661 & 0.0962 & 0.1069 & 0.5356 & 0.1018 & 0.7082 & 0.5844 \\
		AgentReview(GLM) & 0.4700 & 0.4662 & 0.6477 & 0.1224 & 0.2643 & 0.5878 & 0.0729 & 0.6980 & 0.6363 \\
		\midrule
		\multicolumn{10}{@{}l}{\textbf{pointwise - models}} \\
		SEA-E & 0.3707 & 0.3317 & 0.5638 & 0.0508 & 0.1397 & 0.5505 & 0.1218 & 0.6305 & 0.5071 \\
		CycleReviewer-8B & 0.6609 & 0.5338 & 0.6525 & 0.0933 & 0.2775 & 0.5957 & 0.1827 & 0.6956 & 0.6969 \\
		DeepReview-7B & 0.6467 & 0.5412 & 0.5915 & 0.0944 & 0.2971 & 0.6043 & 0.1745 & 0.7190 & 0.6957 \\
		DeepReview-14B & 0.6845 & \underline{0.6254} & \underline{0.6624} & \underline{0.2510} & \underline{0.4014} & \underline{0.6419} & 0.1478 & 0.7204 & \underline{0.8211} \\
		\midrule
		\multicolumn{10}{@{}l}{\textbf{pairwise / listwise}} \\
		NAIP & 0.6025 & 0.5347 & 0.5665 & 0.0695 & 0.1685 & 0.5585 & 0.1379 & 0.6630 & 0.6104 \\
		PairReview(GPT) & 0.6435 & 0.5851 & 0.6130 & 0.1701 & 0.2637 & 0.5925 & 0.2730 & 0.7156 & 0.7387 \\
		PairReview(Gemini) & 0.6246 & 0.5630 & 0.6054 & 0.1261 & 0.2353 & 0.5837 & 0.2920 & 0.7522 & 0.7127 \\
		PairReview(GLM) & 0.6246 & 0.5630 & 0.6325 & 0.1261 & 0.3018 & 0.6066 & 0.3474 & 0.7396 & 0.7499 \\
		\midrule
		{\textbf{CNPE-7B}} & \textbf{0.7192} & \textbf{0.6732} & \textbf{0.7408} & \textbf{0.3464} & \textbf{0.4091} & \textbf{0.6448} & \textbf{0.7076} & \textbf{0.8153} & \textbf{1.0000} \\
		
		\bottomrule
	\end{tabularx}
	\caption{Performance comparison of different methods on ICLR-2025 dataset. Avg. Perf. indicates the average of relative ratios across all metrics, where the maximum value within each metric is normalized to 1. For each metric, \textbf{Best result} and \underline{second-best result} are highlighted.}
	\label{tab:main}
\end{table*}

\section{Experiments}
\label{sec:exp}

\subsection{Experimental Settings}

\paragraph{Dataset Construction}  
Our primary training and test sets are constructed from the ICLR-2025 conference data available at \url{https://openreview.net}. To enable a fair comparison with baseline methods, we follow the train-test split used in DeepReview \cite{zhu2025deepreview}. The dataset includes paper metadata, and the ground truth is defined by the average scores assigned by human reviewers. For the generalization experiments, we use data from five major academic conferences: ICML, NeurIPS, ACL, EMNLP, and NAACL. This extended collection provides grouping information for papers that have been assessed by human experts, capturing a range of quality levels. Additional construction details are provided in Appendix \ref{sec:dataset_construction}.

\paragraph{Basic Configurations}
During training, to avoid potential data leakage from the model inadvertently learning test data, we used Qwen2.5-7B-Instruct \cite{qwen2025qwen25technicalreport} as the base model, and applied LoRA adaptation \cite{hu2022lora} to improve training efficiency. In the reward function, the scaling parameter $\gamma$ was set to 5. For the training dataset, the minimum score difference threshold $d_{\text{min}}$ was 1.5, and the maximum occurrence count $c_{\text{max}}$ was 1 to ensure each paper appeared only once, thereby promoting diversity. During inference, the sample pair fraction parameter $\alpha$ was set to 0.05. The paper acceptance rate was fixed at the average rate of ICLR-2023 and ICLR-2024, 31.4\%. Detailed configurations for training and inference are provided in the Appendix \ref{sec:configs}.

\paragraph{Baseline Methods}
We evaluate three categories of baseline methods: (1) agent-based assessment systems, including AIScientist~\cite{lu2024ai} and AgentReview~\cite{jin2024agentreview}; (2) training-based review models, including SEA~\cite{yu2024automated}, CycleReviewer~\cite{weng2025cycleresearcher}, and DeepReview~\cite{zhu2025deepreview}; and (3) comparison-based evaluation approaches, including NAIP~\cite{zhao2025words} and PairReview~\cite{zhang2025from}. Since AIScientist, AgentReview, and PairReview do not rely on a fixed base model, we mitigate dependence on any single provider by evaluating these methods using LLMs from OpenAI, Google, and Zhipu. Reproducibility details, including the selection of general-purpose LLMs and parameter sizes of fine-tuned models, are provided in Appendix~\ref{sec:baseline_reproduction}.

\paragraph{Evaluation Metrics}
We evaluate our approach using two categories of metrics. The first assesses decision accuracy, formulated as a binary classification task to predict whether a paper should be accepted or rejected. We report Accuracy, F1 score, AUC, and Cohen's $\kappa$. The second evaluates ranking quality, capturing the model's ability to prioritize higher-quality papers over lower-quality ones. Metrics include Spearman's $\rho$, pairwise accuracy, MAP@20, and NDCG@20. Further technical details are provided in Appendix~\ref{sec:evaluation_metrics}.

\subsection{Main Results}

\begin{table*}[t]
	\setlength{\belowcaptionskip}{-5pt} 
	\setlength{\floatsep}{10pt} 
	\setlength{\textfloatsep}{10pt}
	\centering
	\scriptsize
	\setlength{\tabcolsep}{3pt}
	\begin{tabularx}{\textwidth}{@{}
			>{\raggedright\arraybackslash}p{0.14\linewidth}
			*{9}{>{\centering\arraybackslash}X}@{}} 
		\toprule
		& \multicolumn{4}{c}{\textbf{Decision}} & \multicolumn{4}{c}{\textbf{Ranking}} &  \\
		\cmidrule(lr){2-5} \cmidrule(lr){6-9}
		\textbf{Method} & \textbf{Accuracy} & \textbf{F1} & \textbf{AUC} & \textbf{Cohen \(\kappa\)} & \textbf{Spearman \(\rho\)} & \textbf{Pair. Acc.} & \textbf{MAP@20} & \textbf{NDCG@20} & \textbf{Avg. Perf.} \\
		\midrule
		\multicolumn{10}{@{}l}{\textbf{training methods}} \\
		w/o Training & 0.5931 & 0.5263 & 0.5227 & 0.0526 & 0.0841 & 0.5292 & 0.2448 & 0.6847 & 0.5845 \\
		w/o RLVR & \underline{0.6530} & \underline{0.5961} & \underline{0.6566} & \underline{0.1922} & \underline{0.3273} & \underline{0.6159} & 0.2301 & 0.7177 & \underline{0.7744} \\
		w/o SFT cold-start & 0.6325 & 0.5740 & 0.6248 & 0.1480 & 0.2873 & 0.6016 & \underline{0.3315} & \underline{0.7357} & 0.7511 \\
		\midrule
		\multicolumn{10}{@{}l}{\textbf{training data}} \\
		w/o Random (train) & \underline{0.6814} & \underline{0.6291} & \underline{0.6815} & \underline{0.2583} & \underline{0.3598} & \underline{0.6282} & \underline{0.4777} & \underline{0.7807} & \underline{0.8792} \\
		w/o Sim (train) & 0.6640 & 0.6105 & 0.6773 & 0.2211 & 0.3532 & 0.6251 & 0.3697 & 0.7734 & 0.8358 \\
		\midrule
		\multicolumn{10}{@{}l}{\textbf{test data}} \\
		w/o Random (test) & \underline{0.7161} & \underline{0.6695} & 0.7384 & \underline{0.3390} & 0.4011 & 0.6423 & 0.6937 & \underline{0.8069} & \underline{0.9890} \\
		w/o Sim (test) & 0.7098 & 0.6622 & \underline{0.7398} & 0.3244 & \underline{0.4088} & \underline{0.6446} & \underline{0.7019} & 0.7971 & 0.9842 \\
		\midrule
		{\textbf{full}} & \textbf{0.7192} & \textbf{0.6732} & \textbf{0.7408} & \textbf{0.3464} & \textbf{0.4091} & \textbf{0.6448} & \textbf{0.7076} & \textbf{0.8153} & \textbf{1.0000} \\
		\bottomrule
	\end{tabularx}
	\caption{Ablation study results. For each metric, the full model consistently achieves the \textbf{best result}. Also, each group reports its \underline{closest result} to the full model.}
	\label{tab:ablation}
\end{table*}

The main experimental results are summarized in Table~\ref{tab:main}. Our proposed framework consistently outperforms all baselines across both decision and ranking metrics. In particular, compared with the strong baseline DeepReview-14B, it achieves an average improvement of 21.8\% across all metrics.

For the acceptance decision task, our model attains an F1 score of 0.6732 and an AUC of 0.7408, with the latter representing an 11.8\% gain over the closest competitor. This indicates a stronger discriminative ability in distinguishing accepted from rejected papers. For the paper ranking task, the advantages are more pronounced. Our model achieves MAP@20 of 0.7076 and NDCG@20 of 0.8153, with MAP@20 exhibiting a 52.6\% improvement over the second-best model. This substantial gain reflects the inherent advantage of our comparison-native framework, which directly optimizes relative judgments between papers rather than relying on absolute, pointwise scoring. As a result, the model is better aligned with ranking-oriented objectives. Compared with existing pairwise methods, our approach further benefits from task-specific SFT and RL on carefully constructed comparison data, enabling the model to more effectively internalize comparative reasoning patterns. This leads to consistently stronger performance in identifying and prioritizing high-quality papers. 

Remarkably, despite using only 7B parameters, our framework surpasses larger LLMs, including DeepReview-14B, across all metrics, demonstrating both parameter efficiency and the effectiveness of comparison-native learning in our framework. 

\begin{figure*}[htbp]
	\setlength{\belowcaptionskip}{5pt} 
	\setlength{\floatsep}{10pt} 
	\setlength{\textfloatsep}{10pt}
	\centering
	\includegraphics[width=\textwidth]{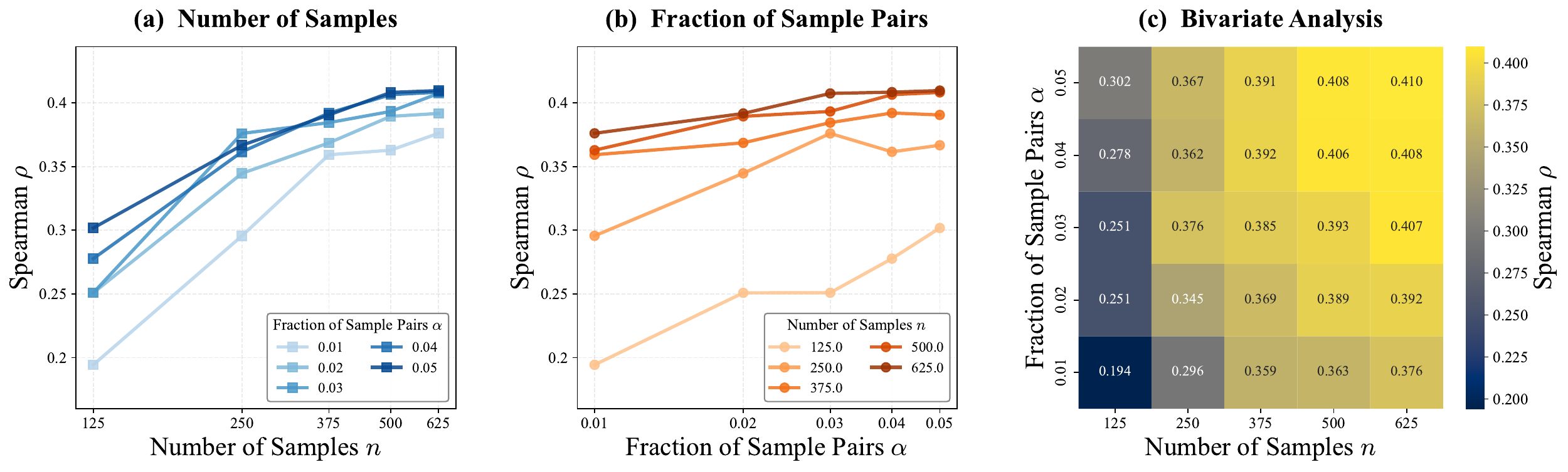}
	\caption{Parameter analysis. Panels (a) and (b) illustrate the relationships of $n$ and $\alpha$ with Spearman's $\rho$ (visualized in logarithmic scale); panel (c) shows their combined effects in a heatmap.}
	\label{fig:parameters}
\end{figure*}

\begin{figure*}[htbp]
	\setlength{\belowcaptionskip}{0pt} 
	\setlength{\floatsep}{10pt} 
	\setlength{\textfloatsep}{10pt}
	\centering
	\includegraphics[width=\textwidth]{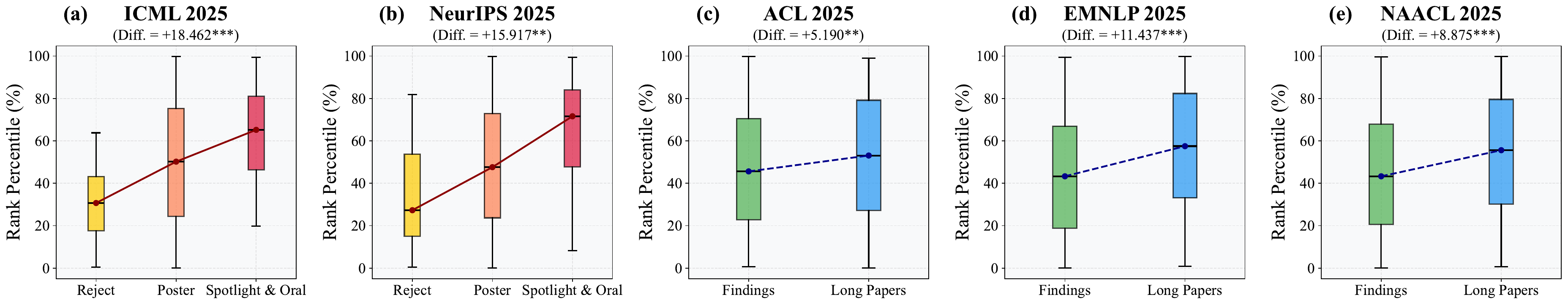}
	\caption{Generalization on previously unseen datasets. The differences between all groups are statistically significant.} 
	\label{fig:generalization}
\end{figure*}

\subsection{Ablation Study}

To assess the impact of training strategies and data construction on the overall performance of the proposed framework, we conducted an extensive ablation study, with results summarized in Table~\ref{tab:ablation}.
We first evaluated different training pipelines. Removing comparison-specific training entirely results in a sharp average performance drop of 41.6\%, indicating that the base model alone is insufficient for learning reliable comparative judgments. Applying SFT without RL also leads to a substantial degradation of 21.6\% on average, while initiating RL directly from the original model similarly produces inferior results. In contrast, the two-stage pipeline, SFT followed by RL, consistently achieves the best performance, showing that the two stages play complementary and necessary roles in learning effective comparison behaviors.

We further investigated the impact of sampling strategies for constructing comparison pairs during training and testing. While similarity-based sampling generally outperforms random sampling when used in isolation, relying on either strategy alone during training leads to clear performance degradation, with average drops of 12.1\% and 16.4\%, respectively. In comparison, excluding either strategy at test time results in only marginal declines (1.1\% and 1.6\%), suggesting that training-time diversity is more critical than test-time diversity. Overall, combining similarity and random-based sampling during training yields the most robust performance, highlighting their complementary contributions to learning relative judgments.  

Finally, we compared alternative algorithms for aggregating preference pairs. Among all candidates, the Bradley-Terry model achieves the highest average performance. Further details are provided in Appendix~\ref{sec:aggregation_algorithms}. 

Overall, the full model consistently achieves the best performance across all metrics, demonstrating that both the comparison-native training paradigm and the carefully designed data construction strategy are essential to the model's effectiveness.
\subsection{Hyperparameter Analysis}

We study the effect of the number of samples at inference ($n$) and the sampling fraction of comparison pairs ($\alpha$) on model performance. To control these parameters, we randomly draw subsets from the full test set, varying both $n$ and $\alpha$, and assess the impact of each $(n, \alpha)$ combination within our framework. For each setting, we conduct 30 independent runs using distinct random seeds. Spearman's $\rho$ is adopted as the primary evaluation metric, and results are reported as the mean across runs. The results are shown in Figure \ref{fig:parameters}.

The results reveal two scaling trends. First, performance improves steadily as the number of comparison samples $n$ increases, indicating that evaluating a larger candidate set provides more reliable relative judgments. Second, increasing the sampling ratio $\alpha$ also yields performance gains, although these improvements are consistently smaller than those obtained by increasing $n$. Together, these findings suggest that model performance can be systematically enhanced by increasing both $n$ and $\alpha$, at the cost of additional computational overhead.

\subsection{Generalization on Unseen Datasets}

We evaluated the model's generalization on unseen papers. For ICML and NeurIPS, we randomly sampled 500 papers per venue and grouped them into Rejected, Poster, and Spotlight \& Oral categories. For ACL, EMNLP, and NAACL, where review outcomes were unavailable, we sampled 250 Long Papers and 250 Findings papers per venue.

Given that percentile rankings are not normally distributed, we employ the non-parametric Mann‑Whitney U test to assess differences between groups (provided in Appendix~\ref{sec:generalization}). As shown in Figure~\ref{fig:generalization}, the score gaps between accepted and rejected papers are substantially larger for ICML (+18.5) and NeurIPS (+15.9) than the differences observed between Findings and Long Papers in ACL (+5.2), EMNLP (+11.4), and NAACL (+8.9). This pattern aligns with the expected magnitude of quality differences across these venues, indicating that the model can capture fine-grained quality distinctions under diverse evaluation settings.

\subsection{Positional Bias Mitigation}
\begin{figure}[htbp]
	\setlength{\belowcaptionskip}{-10pt} 
	\setlength{\floatsep}{10pt} 
	\setlength{\textfloatsep}{10pt}
	\centering
	\includegraphics[width=\columnwidth]{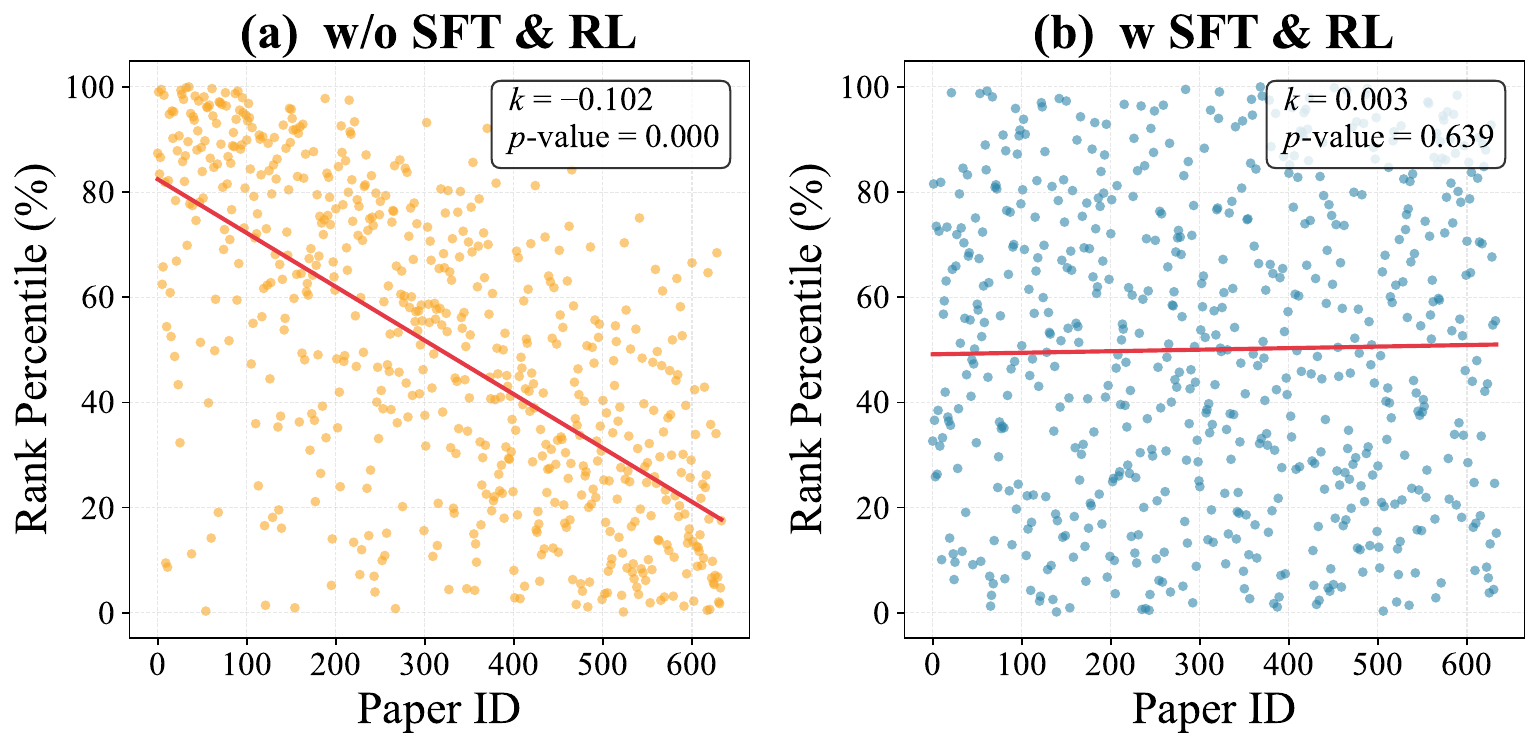}
	\caption{Positional bias. Panel (a) shows the performance of the original Qwen2.5‑7B‑Instruct model; panel (b) shows the performance after SFT and RL.}
	\label{fig:position}
\end{figure}

LLMs are imperfect comparators, as their preference judgments can be distorted by positional bias \cite{wang2024large}. Without appropriate correction, models may systematically favor the first option presented. When such ordered preference pairs are used for ranking, this bias propagates through the Bradley‑Terry aggregation, leading to systematic over- or under-ranking of papers with smaller identifiers and ultimately undermining evaluation reliability. Figure~\ref{fig:position}(a) illustrates this effect: without SFT or RL, the base model exhibits pronounced positional bias, manifested as a clear negative correlation between paper identifiers and ranking percentiles.
This bias can be effectively mitigated through our comparison-native training framework, which combines SFT and RL. After training, the model produces more position-invariant preference judgments, enabling fairer quality comparisons regardless of input order. As shown in Figure~\ref{fig:position}(b), the resulting rankings no longer exhibit a significant correlation between paper identifiers and ranking percentiles, indicating a substantial improvement in evaluation reliability.

\section{Conclusion}
\label{sec:conclusion}

We propose a comparison-native framework that reformulates paper evaluation as collaborative ranking rather than isolated scoring. By integrating graph-based pair sampling with comparison-aware training, the framework consistently improves both ranking and decision performance and generalizes well to unseen venues. These findings show that our framework enables LLMs to learn more accurate and transferable paper comparison capabilities.

\section*{Limitations}

Our work has several limitations across the data aspect, the technical aspect, and the practical aspect. In the data aspect, our experiments rely exclusively on computer science conference papers, which limits the generalizability of our findings. The dataset includes only papers from six leading machine learning and artificial intelligence conferences in 2025. While this narrow time window restricts diversity, it also lowers the risk of information leakage. In the technical aspect, resource constraints restricted us to training a 7B model. Even with best practices applied, this model size cannot match the broader knowledge and deeper language understanding that larger models typically offer. Using only titles and abstracts significantly reduces computational cost and allows wider applicability, but it inevitably constrains information available in full papers, which may cause mild performance degradation. Length limitations on generated reviews may also lead to the loss of fine-grained details. In the practical aspect, although our results represent notable gains over prior work, they still fall short of human reviewing quality. Research on pairwise and listwise approaches remains in its early phase, and our main goal is to demonstrate the promise of this paradigm. With more computational resources, future studies could explore more sophisticated designs, extend to additional fields, and leverage larger models.

\section*{Ethical Considerations}

This work has the potential to improve both the efficiency of paper evaluation and the accessibility of high‑quality feedback. The system may increase fairness across the review process by helping authors without strong peer networks or research resources access early expert guidance, which can raise the quality of submissions and ease the workload of human reviewers. However, the system also introduces important ethical risks. These include the possibility that it could be misused as a replacement for human judgment, that it could reinforce biases present in its training data, that it might further marginalize underrepresented viewpoints, and that it could erode reviewers' skills over time. Because the model is trained largely on top‑tier conference review practices, its implicit standards may disadvantage unconventional research directions, non‑mainstream methods, or communities with limited resources. We stress that although our method advances automated assessment of scientific manuscripts, it is not designed to replace peer review. Its purpose is to support and extend the expertise of human reviewers, and it should only be used as an assistive tool under careful expert supervision. To encourage responsible use, we implemented several safeguards in the system's design and release. These include explicit documentation of system limitations, open‑sourcing the code, and offering practical usage guidelines. We will continue conducting bias audits and standardized evaluations, and we invite the broader community to collaborate on establishing ethical norms for automated reviewing technologies so that these tools can be integrated into the peer‑review ecosystem in a cautious, trustworthy, and beneficial way.

\section*{Acknowledgments}

This work is supported by the National Natural Science Foundation of China (72204087, 72234005), the Chenguang Program of Shanghai Education Development Foundation and Shanghai Municipal Education Commission (23CGA28), the Shanghai Pujiang Program (23PJC030), Young Elite Scientists Sponsorship Program by CAST (YESS20240562). We also appreciate the constructive comments from the anonymous reviewers.

\bibliography{custom}

\appendix

\section{Use of AI Assistants}
\label{sec:use_of_ai}
We used generative artificial intelligence tools solely to assist with language editing, including grammar correction, clarity enhancement, and stylistic consistency. These tools did not generate or contribute to the intellectual, scientific, or substantive content of the manuscript; all factual and conceptual material is entirely the work of the authors. All AI-assisted edits were reviewed, verified, and, where necessary, revised by the authors in accordance with ethical guidelines, ensuring readability improvements without altering the intended meaning. The authors accept full responsibility for the content and conclusions of the work.

\section{Responsible Use}
\label{sec:responsible_use}
While the proposed method offers significant advances in automating scholarly manuscript evaluation, it is not intended to replace human peer review. Its goal is to enhance established practices by integrating algorithmic analysis with the domain expertise, contextual reasoning, and nuanced judgment of qualified reviewers. The system should serve as a computational aid to researchers and the broader academic community, providing alternative perspectives and improving efficiency at selected review stages. Real-world deployment must be guided by rigorous human oversight, and all substantive conclusions should remain under the authority of experienced reviewers to ensure accuracy, fairness, and integrity in scholarly assessment.

\section{Discussion}
\label{sec:discussion}

\subsection{Systemic Perspective of Evaluation}

Our study reflects deeper theoretical foundations and ultimately advocates for a systemic perspective on evaluation. This perspective, we argue, offers a more insightful framework for understanding the nature and implications of our work.

To illustrate, consider a thought experiment in which we are presented with a large set of research papers alongside a hypothetical “technology tree” that will continue to expand over time. In this analogy, the role of paper review is to identify the works most likely to form the critical branches from which the future tree will grow. The decisions made during the review process directly influence the shape and trajectory of this growth. In practice, the task resembles selecting the most promising path among multiple possible futures. Such a task is inherently unsuited to a fixed and absolute standard; rather, it requires comparative judgments that guide the evolution of the tree. Since reviewers are always confronted with new work rather than past work, any model designed to support this process should be trained with a forward-looking objective.

Therefore, using LLMs to fit review scores is inherently unreliable. While it may appear that peer review is simply about predicting an accurate numerical score, its ultimate purpose is to discriminate between the relative quality of contemporary papers, fostering promising directions in scientific inquiry through critical comparisons across parallel research efforts.

This reasoning highlights the need for a systemic and collaborative approach to paper review, as opposed to an atomized and isolated one. From this standpoint, reviewing should be understood as a constructive and interaction-driven process. It does not assume the existence of a single, objective, and universally aligned standard. Instead, promising directions emerge through the interplay of reviewer perspectives and comparative analysis, particularly by examining how new contributions relate to alternative emerging trajectories, as reflected in their connections to other papers.

\subsection{Design Philosophy}

In our comparison-based framework, performance is determined by three components: the paired dataset $\mathcal{D}$, the comparison model $\mathcal{M}$, and the aggregation method $\mathcal{A}$. The evaluation score is $\mathcal{E} = (\mathcal{A} \circ \mathcal{M} \circ \mathcal{D})(\mathcal P)$. Our approach introduces improvements in the following aspects. First, we apply both similarity-based and random sampling strategies to the original dataset to generate more informative paper pairs, denoted as $\mathcal{D}_0 = \mathcal{S}(\mathcal{D}_0)$. Second, we enhance the reasoning and quality-judgment capabilities of the original model through reinforcement learning, denoted as $\mathcal{M}_0 = \mathcal{J}(\mathcal{M}_0)$. This framework leverages the strengths of comparison-based methods by focusing on improving the sampling strategy $\mathcal{S}$ and the judgment capability $\mathcal{J}$, rather than relying on larger datasets $\mathcal{D}_0$ or more powerful models $\mathcal{M}_0$.

\subsection{Paradigm Comparison}

Pointwise and pairwise/listwise methods differ in focus and capability, adopting fundamentally different evaluation strategies. As shown in Table \ref{tab:paradigm}, the distinction is evidenced by the comparison between our model and DeepReview \cite{zhu2025deepreview}. Pointwise methods require detailed analysis of the entire paper and tend to produce evaluations emphasizing textual structure, argumentation, writing quality, and technical accuracy. Pairwise/listwise paradigms model relative relationships among papers within a broader research context, enabling broader and more diversified perspectives. For content evaluation tasks, the two paradigms are complementary: pointwise methods examine the content in depth, while pairwise/listwise methods capture a paper's position, impact, and relative quality in the research landscape. For ranking and recommendation tasks, our approach shows greater potential, as it relies solely on metadata.

\begin{table}[t]
	\centering
	\scriptsize
	\setlength{\tabcolsep}{3pt}
	\begin{tabularx}{0.48\textwidth}{@{}
			>{\raggedright\arraybackslash}p{0.30\linewidth}
			*{2}{>{\centering\arraybackslash}X}@{}} 
		\toprule
		\textbf{Paradigm} & \textbf{Pointwise} & \textbf{Pairwise/Listwise} \\
		\midrule
		\multicolumn{3}{@{}l}{\textbf{Method}} \\
		Representative method & DeepReview & CNPE \\
		\midrule
		\multicolumn{3}{@{}l}{\textbf{Characteristics}} \\
		Text Requirement & Relies on full text & Relies only on metadata \\
		Orientation & Internal & External \\
		Focus & Text features & Domain relationships \\
		Analysis Granularity & Fine, single & Coarse, diverse \\
		Reasoning Length & Long reasoning & Short reasoning \\
		Reasoning Number & Few & Many \\
		\midrule
		\multicolumn{3}{@{}l}{\textbf{Applicable Tasks}} \\
		Acceptance decision & \checkmark\ Good & \checkmark\ Best \\
		Paper ranking & \checkmark\ Good & \checkmark\ Best \\
		Review generation & \checkmark\ Good & \checkmark\ Good \\
		Paper recommendation & $\times$ Not applicable & \checkmark\ Best \\
		\bottomrule
	\end{tabularx}
	\caption{Comparison between pointwise and pairwise/listwise paradigms.}
	\label{tab:paradigm}
\end{table}

\subsection{Methodology Comparison}

To highlight the distinctions between our proposed framework and existing methods, we conducted a systematic comparative analysis, with the key theoretical differences summarized in Table~\ref{tab:model_comparison}. Specifically:  (1) conventional approaches primarily rely on direct scoring \cite{zhu2025deepreview}; (2) several early studies implement comparison only partially rather than throughout both training and inference \cite{zhao2025words,zhang2025from}; (3) our framework maintains a consistent comparison-based approach from training to inference, which leads to superior performance.

\begin{table}[t]
	\centering
	\scriptsize
	\setlength{\tabcolsep}{3pt}
	\begin{tabularx}{\linewidth}{@{}
			>{\raggedright\arraybackslash}p{0.16\linewidth}
			*{2}{>{\centering\arraybackslash}X}
			>{\raggedright\arraybackslash}p{0.22\linewidth}
			>{\raggedright\arraybackslash}p{0.36\linewidth}@{}} 
		\toprule
		\textbf{Model} & \textbf{CbT} & \textbf{CbI} & \textbf{Philosophy} & \textbf{Details} \\
		\midrule
		DeepReview  \newline \cite{zhu2025deepreview} &  &  & From \newline Scoring, \newline For \newline Scoring & A standard end-to-end scoring methodology that leverages LLMs to assign scores directly from textual input. \\
		NAIP  \newline \cite{zhao2025words} & \checkmark &  & From \newline Comparison, \newline For \newline Scoring & Incorporates listwise comparison data in trainingy, but reverts to an isolated, opaque absolute score during inference. \\
		PairReview  \newline \cite{zhang2025from} &  & \checkmark & From \newline Common Sense, \newline For \newline Comparison & Uses pairwise preference modeling but relies on general LLMs lacking domain training, causing notable position bias. \\
		\textbf{CNPE} & \checkmark & \checkmark & From \newline Comparison, \newline For \newline Comparison & A native comparison-based review framework embeds comparison principles to deliver more reliable, informative evaluations. \\
		\bottomrule
	\end{tabularx}
	\caption{Comparison of different review models in terms of training and inference paradigms. CbT: Comparison-based Training, CbI: Comparison-based Inference}
	\label{tab:model_comparison}
\end{table}

\subsection{Application Prospects}

The proposed comparison framework offers versatile applicability beyond controlled experimental settings. In the following, we outline its potential applications in three representative contexts.

\paragraph{Evaluating New Research}

Our method is well suited for evaluating recently published papers, as it partially resolves a key limitation of existing score-based approaches: the weak generalization of models trained on historical data when assessing emerging research. Conventional evaluation often applies outdated criteria, leading to misalignment with evolving standards of scientific quality. In contrast, our pairwise comparison framework assesses papers based on relative quality rather than absolute scores tied to obsolete knowledge. This design reduces the impact of shifting benchmarks and improves generalization. Experiments confirm that our method generalizes effectively across diverse conference evaluation tasks, demonstrating its suitability for assessing new submissions.

\paragraph{Enhancing Academic Recommendation}

The proposed approach is also highly applicable to academic recommendation systems, where quality is critical in addition to topical relevance. Our framework identifies innovative and high-quality contributions, achieving substantial gains over baselines on ranking metrics such as MAP and NDCG. It shows strong alignment with human recommendations, particularly in highlighting groundbreaking papers in frontier domains. Importantly, the model only uses titles and abstracts, which alleviates barriers arising from restricted full-text access. This feature is especially beneficial in non-computer-science fields, enhancing early-stage automated recommendation and quality filtering, and increasing the feasibility of real-world deployment.

\paragraph{Guiding Scholarly Progress}

Our results point to a promising route for advancing science by fostering constructive competition and mapping potential trajectories of technological development. While fine-grained peer review of full texts remains essential, it does not fully capture broader patterns of scientific progress. We advocate the integration of complementary mechanisms capable of identifying and guiding emerging research directions. By systematically comparing multiple research pathways, such methods can inform the evolution of influential scientific paradigms. An effective review framework should combine two perspectives: evaluation of writing quality, originality, and theoretical soundness, and assessment of a paper's potential significance and representativeness in shaping future advances. Such comparative evaluation enables a holistic view of scientific evolution, which forms a central focus of our future work.

\section{Methodology Details}

\subsection{Data Sampling}
\label{sec:gbr_br_alg}

In the experiments, the parameters $k_{\mathrm{e}}$ and $k_{\mathrm{r}}$ are set to 50 and 25, respectively. $\mathrm{Embed}$ employs the Qwen3-embedding-0.6B model, while $\mathrm{Rerank}$ applies the Qwen3-reranker-0.6B model.

\subsection{Training}
\label{sec:rl_alg}

The following section presents details of our supervised fine-tuning and reinforcement learning procedures during training.

\paragraph{SFT Cold Start}

During the supervised fine-tuning stage, for synthesizing reasoning-chain data, we use the instruct model Qwen3-235B-A22B-Instruct-2507-AWQ. This model has a large number of parameters and can generate reliable reasoning text. Due to limited computational resources, we adopt short reasoning chains and restrict the total length of each reasoning sequence to approximately 200 words.

\paragraph{Reinforcement Learning}

Our reinforcement learning procedure builds upon a refined GRPO algorithm \cite{shao2024deepseekmath} with several modifications \cite{yu2025dapo,liu2025understanding}.

We avoid normalizing the group standard deviation to eliminate length bias arising from varying problem difficulty. The upper clipping threshold in the loss is set slightly above the lower bound to promote policy exploration. In practice, the parameters $\varepsilon_{\mathrm{low}}$ and $\varepsilon_{\mathrm{high}}$ are set to the commonly used values of 0.2 and 0.28, respectively. The KL loss constraint is removed to enable more flexible policy updates. The optimization objective is:

\section{Experiment Details}

\subsection{Dataset Construction}
\label{sec:dataset_construction}

To ensure fair comparison across review paradigms and to prevent future-data leakage, we adopted the following design principles in constructing the dataset.

\paragraph{Ensuring Fair Comparison}
Differences in training data can substantially affect the evaluation of LLM-based review systems. To remove this confounding factor and ensure a fair comparison with leading baselines, we use exactly the same training and test sets as the strongest existing pointwise model, DeepReview \cite{zhu2025deepreview}. Our training corpus is a strict subset of DeepReview-13k, and evaluation is conducted on the established DeepReviewBench benchmark rather than reconstructing datasets from ICLR. This design allows a direct comparison of performance between pointwise and pairwise/listwise paradigms while minimizing bias from dataset discrepancies.

\paragraph{Preventing Information Leakage}
We take strict measures to avoid information leakage in the test sets. Since 2025 data occur chronologically after 2024, including them in training could inadvertently introduce future knowledge and artificially boost performance on 2024 test sets. Our goal is to evaluate models on reviewing newly submitted papers, making it essential that test data do not precede the training set. Therefore, all test data in this study are drawn from 2025 across ICLR, ICML, NeurIPS, ACL, EMNLP, and NAACL, with publication dates strictly later than those in the training corpus. This setup ensures that the evaluation remains unaffected by future-data leakage.

\subsection{Basic Configurations}
\label{sec:configs}
The following content presents details of our model usage and parameter configuration in both training and inference.

\paragraph{Training Configurations}
To prevent potential data leakage arising from the model inadvertently learning from test data, we adopted Qwen2.5‑7B‑Instruct \cite{qwen2025qwen25technicalreport} as the base model. The main training process was conducted on two RTX 6000 Pro GPUs with 96 GB memory each. LoRA adaptation \cite{hu2022lora} was applied with rank 16 and LoRA‑alpha 32 to enhance training efficiency. Short reasoning chains were employed, and the output length was limited to 512 tokens to improve efficiency and reduce token consumption. For samples exceeding the predefined context length, a truncation strategy was used. In the reward function, the scaling parameter $\gamma$ was set to 5. Supervised fine‑tuning was performed for one epoch with a learning rate of 5e‑4, an original batch size of 2, and gradient accumulation over 16 steps. Reinforcement learning was performed for one epoch with a learning rate of 5e‑5, an original batch size of 4, and gradient accumulation over 16 steps, generating eight trajectories per instance via a single rollout. Within the training dataset, the minimum score difference threshold $d_{\text{min}}$ was set to 1.5, and the maximum occurrence count $c_{\text{max}}$ was set to 1, ensuring that each paper appeared only once to promote diversity.

\paragraph{Inference Configurations}
The sample‑pair fraction parameter $\alpha$ was set to 0.05, and $n$ was always equal to the actual size of the test set. For ICLR‑2025, $n$ is equal to 634, as the test set was aligned with that used in DeepReview. For the other five conferences, $n$ is equal to 500, corresponding to the number of samples selected for the respective test sets. The paper acceptance rate was fixed to the average acceptance rate of ICLR‑2023 and ICLR‑2024, which is 31.4\%.

\subsection{Baseline Reproduction}
\label{sec:baseline_reproduction}

We successfully reproduced three major categories of baseline models in our experiments: (1) Agent-based assessment systems, including reproductions of AIScientist~\cite{lu2024ai} and AgentReview~\cite{jin2024agentreview}.
(2) Training-based review models, with reproductions of SEA~\cite{yu2024automated} (restricted to the SEA-E variant because SEA-EA is not available), CycleReview~\cite{weng2025cycleresearcher} (8B version only due to resource constraints), and DeepReview~\cite{zhu2025deepreview} (7B and 14B versions).
(3) Comparison-based evaluation approaches, including reproductions of PairReview~\cite{zhang2025from} and NAIP~\cite{zhao2025words}.

For AIScientist~\cite{lu2024ai}, AgentReview~\cite{jin2024agentreview}, and PairReview~\cite{zhang2025from}, we intentionally incorporated multiple LLMs developed by different institutions to mitigate the risk of drawing conclusions overly reliant on models from a single vendor. All methods do not rule out the possibility of data leakage, so the metrics may be inflated. Table~\ref{tab:llm_info} details the specific models used.

\begin{table}[t]
	\centering
	\scriptsize
	\setlength{\tabcolsep}{3pt}
	\begin{tabularx}{\columnwidth}{@{}
			>{\centering\arraybackslash}p{0.18\linewidth}
			>{\centering\arraybackslash}p{0.16\linewidth}
			>{\centering\arraybackslash}p{0.30\linewidth}
			>{\centering\arraybackslash}X@{}}
		\toprule
		\textbf{Abbreviation} & \textbf{Company} & \textbf{Model Name} & \textbf{Max Context Length} \\
		\midrule
		GPT & OpenAI & GPT-oss-120B & 128k \\
		Gemini & Google & Gemini-2.5-Flash-Lite & 1M \\
		GLM & Zhipu & GLM-4.5-Air & 128k \\
		\bottomrule
	\end{tabularx}
	\caption{LLMs used in our reproduction experiments.}
	\label{tab:llm_info}
\end{table}

Overall, our reproduction adheres to the methodological framework proposed by Zhu~\cite{zhu2025deepreview}, successfully covering all baselines described in that work while extending the set with additional models. Several baselines could not be fully reproduced, for the following reasons: (1) AgentReviewers~\cite{lu2025agent}: The code and model remain unreleased as of November 2025, making reproduction infeasible. (2) ReviewRL~\cite{zeng2025reviewrl}: Its training set overlaps with our test set, raising concerns about data leakage. Retraining the model from scratch would require computational resources beyond our budget. (3) TreeReview~\cite{chang2025treereview}: Although executable, the method processes each paper in roughly thirty minutes and requires dynamic loading of an open-source model, which severely limits parallelization. Running the full test set would take more than ten days, rendering it impractical.

\subsection{Evaluation Metrics}
\label{sec:evaluation_metrics}

The following content presents the complete definitions of the evaluation metrics, provides their formal mathematical descriptions, and additionally reports experimental results for metrics not included in the main text.

\paragraph{Metric Definitions}
We assess our model's performance using two metric categories, as summarized in Table~\ref{tab:main_metrics}. The first category measures decision accuracy, framed as a binary classification task to predict whether a paper should be accepted or rejected. We report Accuracy, F1 score (harmonic mean of precision and recall, suitable for class-imbalanced settings), AUC (Area Under the ROC Curve), and Cohen's $\kappa$ (agreement beyond chance). The second category measures ranking quality, reflecting the ability to position higher-quality papers ahead of lower-quality ones while maintaining their relative order. We adopt Spearman's $\rho$ (rank correlation), Pairwise Accuracy, MAP@k (Mean Average Precision at k, with relevance defined above the acceptance threshold in our research), and NDCG@k (Normalized Discounted Cumulative Gain, accounting for item position and relevance).

\begin{table}[t]
	\centering
	\scriptsize
	\setlength{\tabcolsep}{3pt}
	\renewcommand{\arraystretch}{1.0}
	\begin{tabularx}{\columnwidth}{@{}
			>{\raggedright\arraybackslash}p{0.18\linewidth}
			>{\raggedright\arraybackslash}X@{}}
		\toprule
		\textbf{Metrics} & \textbf{Formula} \\
		\midrule
	\end{tabularx} \\ [-0.4em]
	\renewcommand{\arraystretch}{2}
	\begin{tabularx}{\columnwidth}{@{}
			>{\raggedright\arraybackslash}p{0.18\linewidth}
			>{\raggedright\arraybackslash}X@{}}
		Accuracy & \tiny$\displaystyle\mathrm{Accuracy} = \frac{1}{N} \sum_{i=1}^{N} \mathbb{I}(y_i = \hat{y}_i)$ \\
		F1 & \tiny$\displaystyle\mathrm{F1} = 2 \cdot \frac{\mathrm{Precision} \cdot \mathrm{Recall}}{\mathrm{Precision} + \mathrm{Recall}}$ \\
		AUC & \tiny$\displaystyle\mathrm{AUC} = P(\hat{s}_i > \hat{s}_j \mid y_i = 1, y_j = 0)$ \\
		Cohen \tiny$\displaystyle\kappa$ & $\mathrm{\kappa} = \frac{p_o - p_e}{1 - p_e}$ \\
		Spearman \tiny$\displaystyle\rho$ & $\mathrm{\rho} = 1 - \frac{6 \sum_{i=1}^{N} d_i^2}{N(N^2 - 1)}$ \\
		Pair. Acc. & \tiny$\displaystyle\mathrm{Pair. Acc.} = \frac{1}{M} \sum_{(i,j) \in \mathcal{P}} \mathbb{I}\big( (y_i - y_j)(\hat{y}_i - \hat{y}_j) > 0 \big)$ \\
		MAP@k & \tiny$\displaystyle\mathrm{MAP@k} = \frac{1}{\min(R, k)} \sum_{k=1}^{k} P(k) \cdot \mathrm{rel}(k)$ \\
		NDCG@k & \tiny$\displaystyle\mathrm{NDCG@k} = \frac{\mathrm{DCG@k}}{\mathrm{IDCG@k}}$ \\
		\bottomrule
	\end{tabularx}
	\caption{Main metrics.}
	\label{tab:main_metrics}
\end{table}

\begin{table}[t]
	\centering
	\scriptsize
	\setlength{\tabcolsep}{3pt}
	\renewcommand{\arraystretch}{1.0}
	\begin{tabularx}{\columnwidth}{@{}
			>{\raggedright\arraybackslash}p{0.18\linewidth}
			>{\raggedright\arraybackslash}X@{}}
		\toprule
		\textbf{Metrics} & \textbf{Formula} \\
		\midrule
	\end{tabularx} \\ [-0.4em]
	\renewcommand{\arraystretch}{2.5}
	\begin{tabularx}{\columnwidth}{@{}
			>{\raggedright\arraybackslash}p{0.18\linewidth}
			>{\raggedright\arraybackslash}X@{}}
		Jaccard & {\tiny$\displaystyle\mathrm{Jaccard} = \frac{TP}{TP + FP + FN}$} \\
		F1-weighted & {\tiny$\displaystyle\mathrm{F1_{weighted}} = \sum_c \frac{N_c}{N} \cdot \mathrm{F1}_c$} \\
		Kendall $\tau$ & {\tiny$\displaystyle\mathrm{\tau} = \frac{C - D}{N(N-1)/2}$} \\
		\bottomrule
	\end{tabularx}
	\caption{Additional metrics.}
	\label{tab:other_metrics}
\end{table}

\begin{table*}[t]
	\centering
	\scriptsize
	\setlength{\tabcolsep}{3pt}
	\begin{tabularx}{\textwidth}{@{}
			>{\raggedright\arraybackslash}p{0.14\linewidth}
			*{9}{>{\centering\arraybackslash}X}@{}} 
		\toprule
		& \multicolumn{2}{c}{\textbf{Decision}} & \multicolumn{7}{c}{\textbf{Ranking}} \\
		\cmidrule(lr){2-3} \cmidrule(lr){4-10}
		\textbf{Method} & \textbf{Jaccard} & \textbf{F1-weighted} & \textbf{Kendall $\tau$} & \textbf{MAP@10} & \textbf{NDCG@10} & \textbf{MAP@50} & \textbf{NDCG@50} & \textbf{MAP@all} & \textbf{NDCG@all} \\
		\midrule
		\multicolumn{10}{@{}l}{\textbf{pointwise - agents}} \\
		AIScientist(GPT) & 0.1429 & 0.6362 & 0.1924 & \underline{0.5975} & \underline{0.7459} & \underline{0.3684} & \underline{0.7709} & \underline{0.4574} & \underline{0.9594} \\
		AIScientist(Gemini) & 0.2817 & 0.5776 & 0.1060 & 0.0917 & 0.5836 & 0.1188 & 0.6863 & 0.3528 & 0.9462 \\
		AIScientist(GLM) & 0.3224 & 0.3001 & 0.1685 & 0.1608 & 0.6863 & 0.1966 & 0.7550 & 0.3997 & 0.9558 \\
		AgentReview(GPT) & 0.2811 & 0.5236 & 0.0318 & 0.2611 & 0.6324 & 0.1839 & 0.6883 & 0.3520 & 0.9448 \\
		AgentReview(Gemini) & 0.2906 & 0.5544 & 0.0688 & 0.1400 & 0.6980 & 0.1079 & 0.7109 & 0.3449 & 0.9519 \\
		AgentReview(GLM) & \underline{0.3438} & 0.4489 & 0.1579 & 0.0500 & 0.6895 & 0.1508 & 0.7356 & 0.3920 & 0.9546 \\
		\midrule
		\multicolumn{10}{@{}l}{\textbf{pointwise - models}} \\
		SEA-E & 0.3272 & 0.2707 & 0.0880 & 0.1133 & 0.5966 & 0.1162 & 0.6608 & 0.3426 & 0.9434 \\
		CycleReview-7B & 0.1699 & 0.6259 & 0.1844 & 0.2344 & 0.6602 & 0.1828 & 0.7016 & 0.3989 & 0.9524 \\
		DeepReview-7B & 0.1913 & 0.6245 & 0.1971 & 0.2153 & 0.6858 & 0.1922 & 0.7418 & 0.3893 & 0.9556 \\
		DeepReview-14B & 0.3127 & \underline{0.6817} & \underline{0.2719} & 0.1222 & 0.6775 & 0.2118 & 0.7424 & 0.4267 & 0.9578 \\
		\midrule
		\multicolumn{10}{@{}l}{\textbf{pairwise / listwise}} \\
		NAIP & 0.2174 & 0.6020 & 0.1126 & 0.1633 & 0.6578 & 0.1695 & 0.6948 & 0.3814 & 0.9483 \\
		PairReview(GPT) & 0.2733 & 0.6440 & 0.1783 & 0.4175 & 0.7247 & 0.1716 & 0.7209 & 0.3852 & 0.9559 \\
		PairReview(Gemini) & 0.2492 & 0.6251 & 0.1612 & 0.3267 & 0.7212 & 0.2963 & 0.7611 & 0.4069 & 0.9580 \\
		PairReview(GLM) & 0.2492 & 0.6251 & 0.2054 & 0.5014 & 0.7233 & 0.2591 & 0.7542 & 0.3968 & 0.9569 \\
		\midrule
		\textbf{CNPE-7B} & \textbf{0.3798} & \textbf{0.7196} & \textbf{0.2789} & \textbf{0.6818} & \textbf{0.7953} & \textbf{0.5719} & \textbf{0.8084} & \textbf{0.5455} & \textbf{0.9675} \\
		\bottomrule
	\end{tabularx}
	\caption{Performance comparison on other metrics. For each metric, \textbf{Best result} and \underline{second-best result} are highlighted.}
	\label{tab:main2}
\end{table*}

To validate robustness, we further report complementary metrics beyond the primary ones, as detailed in Table~\ref{tab:other_metrics}: for decision accuracy,  we report the Jaccard coefficient (overlap ratio between predicted and ground-truth label sets), F1-weighted (per-class F1 weighted by sample proportions); and for ranking quality, we report Kendall's $\tau$ (rank concordance).

\paragraph{Rationale of Metric Selection}
Absolute scores vary substantially across journals and even across publication years within the same journal, which limits their applicability across datasets. Ranking accuracy, by contrast, provides a more intrinsic assessment of a paper's quality. Although many prior studies report absolute-score metrics, we do not consider them reliable and refrain from converting rankings into absolute scores. Consequently, such absolute-score metrics are excluded from this study.

\paragraph{Results on Additional Metrics}

We further evaluated our approach using several alternative metrics that remain within the two categories described above, namely ranking-based and accuracy-based measures. The results show that our method consistently attains the highest performance across all metrics, indicating that its effectiveness is robust under diverse evaluation criteria.
See Table \ref{tab:main2} for all results.

\begin{table*}[t]
	\centering
	\scriptsize
	\setlength{\tabcolsep}{3pt}
	\begin{tabularx}{\textwidth}{@{}
			>{\raggedright\arraybackslash}p{0.16\linewidth}
			*{9}{>{\centering\arraybackslash}X}@{}} 
		\toprule
		& \multicolumn{4}{c}{\textbf{Decision}} & \multicolumn{4}{c}{\textbf{Ranking}} &  \\
		\cmidrule(lr){2-5} \cmidrule(lr){6-9}
		\textbf{Method} & \textbf{Accuracy} & \textbf{F1} & \textbf{AUC} & \textbf{Cohen \(\kappa\)} & \textbf{Spearman \(\rho\)} & \textbf{Pair. Acc.} & \textbf{MAP@20} & \textbf{NDCG@20} & \textbf{Avg. Perf.} \\
		\midrule
		\multicolumn{10}{@{}l}{\textbf{graph-based algorithms}} \\
		Eigenvector Centrality & 0.7066 & 0.6585 & 0.7381 & 0.3170 & 0.4071 & 0.6442 & 0.4581 & 0.7702 & 0.9306 \\
		Katz Centrality & 0.7066 & 0.6585 & 0.7382 & 0.3170 & 0.4096 & 0.6449 & 0.4564 & 0.7679 & 0.9308 \\
		HITS & 0.6782 & 0.6255 & 0.7022 & 0.2509 & 0.3642 & 0.6283 & 0.3919 & 0.7339 & 0.8563 \\
		PageRank & \underline{0.7161} & \underline{0.6695} & 0.7385 & \underline{0.3390} & 0.4093 & 0.6448 & 0.5412 & 0.7934 & 0.9613 \\
		\midrule
		\multicolumn{10}{@{}l}{\textbf{probabilistic comparison models}} \\
		Thurstone-Mosteller & \underline{0.7161} & \underline{0.6695} & 0.7405 & \underline{0.3390} & 0.4092 & 0.6449 & 0.6055 & 0.7988 & 0.9738 \\
		Bradley-Terry (MAP) & \textbf{0.7192} & \textbf{0.6732} & \textbf{0.7439} & \textbf{0.3464} & \textbf{0.4124} & \textbf{0.6459} & 0.5861 & 0.7996 & 0.9761 \\
		Bradley-Terry (MCMC) & \textbf{0.7192} & \textbf{0.6732} & \underline{0.7434} & \textbf{0.3464} & \underline{0.4118} & \underline{0.6458} & \underline{0.6310} & \underline{0.8083} & \underline{0.9851} \\
		Bradley-Terry (MLE)  & \textbf{0.7192} & \textbf{0.6732} & 0.7408 & \textbf{0.3464} & 0.4091 & 0.6448 & \textbf{0.7076} & \textbf{0.8153} & \textbf{0.9983} \\
		\bottomrule
	\end{tabularx}
	\caption{Performance comparison of different aggregation methods. For Katz Centrality, we set $\alpha$ to 0.1; for PageRank, we set $\alpha$ to 0.85. These hyperparameters have been carefully tuned. For each metric, \textbf{Best result} and \underline{second-best result} are highlighted.}
	\label{tab:agg_methods}
\end{table*}

\section{Aggregation Algorithms}
\label{sec:aggregation_algorithms}

To understand how different aggregation strategies might influence the final ranking outcomes, we explored several alternatives to the Bradley‑Terry model for pairwise preference modeling:
These alternatives can be broadly grouped into two categories: (1) The first group comprises graph-based approaches, which leverage the structural properties of a directed graph to infer relative paper quality. In our construction, each directed edge encodes the preference of the LLM between a given pair of papers. The graph representation then enables the application of centrality-based ranking techniques. Representative methods we evaluated include Eigenvector Centrality, Katz Centrality, HITS, and PageRank. 
(2) The second group is based on classical probabilistic modeling. These methods share conceptual similarities with Bradley‑Terry but introduce variations such as Gaussian distribution assumptions or Bayesian sampling. Our evaluation includes the Thurstone‑Mosteller model estimated via MLE, as well as two bayesian Bradley‑Terry approaches using MAP and MCMC.

Table \ref{tab:agg_methods} shows the results under different aggregation methods. Empirically, we find that the plain Bradley‑Terry model with maximum likelihood estimation already delivers the strongest overall performance, so we select this method as the preference aggregation approach.

\section{Generalization}
\label{sec:generalization}

\begin{table*}[t]
	\centering
	\scriptsize
	\setlength{\tabcolsep}{3pt}
	\begin{tabularx}{\textwidth}{@{}
			>{\raggedright\arraybackslash}p{0.13\linewidth}
			>{\raggedright\arraybackslash}p{0.12\linewidth}
			>{\raggedright\arraybackslash}p{0.12\linewidth}
			>{\centering\arraybackslash}p{0.07\linewidth}
			>{\centering\arraybackslash}p{0.07\linewidth}
			>{\centering\arraybackslash}p{0.07\linewidth}
			>{\centering\arraybackslash}p{0.07\linewidth}
			>{\centering\arraybackslash}p{0.07\linewidth}
			>{\centering\arraybackslash}p{0.09\linewidth}
			>{\centering\arraybackslash}p{0.06\linewidth}@{}}
		\toprule
		\textbf{Venue} & \textbf{Group1} & \textbf{Group2} & \textbf{Median1} & \textbf{Median2} & \textbf{Diff.} & \textbf{$U$} & \textbf{$Z$} & \textbf{$p$-value} & \textbf{Sig.} \\
		\midrule
		ICML 2025 & Reject & Accept & 30.7 & 51.7 & +21.0 & 4168.0 & -3.285 & 0.001 & *** \\
		& Reject & Poster & 30.7 & 50.2 & +19.5 & 4008.0 & -3.034 & 0.002 & *** \\
		& Reject & Spotlight \& Oral & 30.7 & 65.2 & +34.5 & 160.0 & -4.743 & 0.000 & *** \\
		& Poster & Spotlight \& Oral & 50.2 & 65.2 & +15.0 & 5801.0 & -2.820 & 0.005 & *** \\
		NeurIPS 2025 & Reject & Accept & 27.3 & 50.7 & +23.4 & 3272.0 & -2.414 & 0.016 & ** \\
		& Reject & Poster & 27.3 & 47.6 & +20.3 & 3026.0 & -2.100 & 0.036 & ** \\
		& Reject & Spotlight \& Oral & 27.3 & 71.6 & +44.3 & 246.0 & -3.987 & 0.000 & *** \\
		& Poster & Spotlight \& Oral & 47.6 & 71.6 & +24.0 & 8717.0 & -4.014 & 0.000 & *** \\
		ACL 2025     & Findings & Long Papers & 45.6 & 53.1 & +7.5  & 28006.0 & -2.008 & 0.045 & ** \\
		EMNLP 2025   & Findings & Long Papers & 43.3 & 57.5 & +14.2 & 24102.0 & -4.425 & 0.000 & *** \\
		NAACL 2025   & Findings & Long Papers & 43.3 & 55.6 & +12.3 & 25703.0 & -3.434 & 0.001 & *** \\
		\bottomrule
	\end{tabularx}
	\caption{Mann-Whitney U test summary ($^{***}p < 0.01$, $^{**}p < 0.05$, $^{*}p < 0.1$, $ns$: not significant).}
	\label{tab:mann_whitney_summary}
\end{table*}

As shown in Table \ref{tab:mann_whitney_summary}, the detailed results from the Mann-Whitney U test are as follows. In ICML 2025 ($p < 0.01$) and NeurIPS 2025 ($p < 0.05$), there are significant differences in the median percentile rankings between the accept and reject groups. Significant differences (all $p < 0.05$) exist between all pairs of different groups, indicating that the model can clearly distinguish between acceptance levels such as Reject, Poster, and Spotlight \& Oral. For ACL, EMNLP, and NAACL, due to the lack of detailed review results, we compared the Findings and Long Papers categories. The results show that in ACL 2025 ($p < 0.05$), EMNLP 2025 ($p < 0.01$), and NAACL 2025 ($p < 0.01$), the median percentile rankings of Long Papers are significantly higher than those of Findings. Although both types of papers went through peer review, the model was still able to capture the relatively higher quality signals of Long Papers, and the differences were statistically significant. In terms of specific rankings, the difference in median percentile rankings between rejected and accepted papers in ICML was +21.0, and in NeurIPS it was +23.4; in contrast, the differences between Findings and Long Papers in ACL, EMNLP, and NAACL were +7.5, +14.2, and +12.3, respectively. It can be seen that the absolute differences between accepted and rejected papers in ICML and NeurIPS (21.0‑23.4) are significantly larger than the differences between Findings and Long Papers (7.5‑14.2), indicating that the former reflects a greater quality gap. This perfectly matches real-world conditions and shows that the model has the ability to distinguish fine-grained quality differences. 
The above results demonstrate that the differences among all groups are statistically significant. This indicates that the model possesses a systematic ability to distinguish the quality of unseen papers, thereby exhibiting strong generalization capability.

\section{Prompt Design}
\label{sec:prompt_design}

\subsection{Prompts for Comparative Evaluation}

\begin{tcolorbox}[
	colback=blue!5, 
	colframe=blue!50, 
	title=Prompt,
	fontupper=\scriptsize,
	fonttitle=\bfseries\small, 
	left=2mm, right=2mm, 
	top=2mm, bottom=2mm,  
	boxsep=0mm, 
	toptitle=1mm, bottomtitle=1mm,
	before upper={\linespread{1.25}\selectfont}
	]
	Your response must be about 200 words in length.  \\
	Please act as an impartial judge and evaluate the quality of the following two papers. As the area chair for a top ML conference, you can only select one paper. Start with a brief meta-review / reasoning of the pros and cons for each paper (two sentences), and then provide your choice in a binary format. Start with a brief meta-review / reasoning of the pros and cons for each paper, focusing on novelty, significance, clarity, methodology, and practical implications. Be very critical and do not be biased by what the author claimed. Finally, provide your choice in a binary format. \\
	
	Please provide your analysis in JSON format. \\
	
	\#\#\# Paper 1: \\
	Submission Title: \{title\_1\} \\
	``` \\
	Abstract: \{abstract\_1\} \\
	``` \\
	
	\#\#\# Paper 2: \\
	Submission Title: \{title\_2\} \\
	``` \\
	Abstract: \{abstract\_2\} \\
	``` \\
	
	Your JSON output should look like this: \\
	\{\{ \\
	\hspace*{1em} "paper\_1\_review": "Your meta-review and reasoning for paper 1", \\
	\hspace*{1em} "paper\_2\_review": "Your meta-review and reasoning for paper 2", \\
	\hspace*{1em} "chosen\_paper": "paper\_1 or paper\_2" \\
	\}\}
	\label{prompt:prompt1}
\end{tcolorbox}

As shown above, our prompts for comparative evaluation follow a design approach similar to that of PairReview \citep{zhang2025from}. The main enhancement is the incorporation of explicit constraints on review length, thereby encouraging the model to reach a conclusion using minimal reasoning steps. A further distinction is that, in designing these comparison prompts, we restrict the input to only the titles and abstracts of the paper pairs under comparison. These sections provide high information density and sufficient context to support reliable comparative judgments \cite{zhou2024llm,hopner2025automatic,zhao2025words}.

\subsection{Prompt for Comprehensive Review}

\begin{tcolorbox}[
	colback=red!5, 
	colframe=red!50, 
	title=Prompt,
	fontupper=\scriptsize,
	fonttitle=\bfseries\small, 
	left=2mm, right=2mm, 
	top=2mm, bottom=2mm,  
	boxsep=0mm, 
	toptitle=1mm, bottomtitle=1mm,
	before upper={\linespread{1.25}\selectfont}
	]
	Please output the paper review in two paragraphs, strictly following the content requirements and do not add any other information: \\
	1. Summarize the core content of the review of THIS PAPER. \\
	2. Compare this paper with other papers in a comparative review, using a longer description to make a detailed comparison with other important and similar papers. \\
	
	\#\# THIS PAPER \\
	``` \\
	{str(answers.focal\_papers)} \\
	``` \\
	
	\#\# other papers \\
	``` \\
	{str(answers.other\_papers)} \\
	```
\end{tcolorbox}

As shown above, the initial outputs from LLMs in our setting are tailored for comparative analysis and do not provide explicit or definitive conclusions. Instead, conclusions emerge through synthesis across multiple evaluative dimensions. To this end, we introduce supplementary prompts that leverage the GLM-4.5-Air model to integrate all individual review remarks and distill them into a finalized research assessment. In this study, direct comparison between the generated reviews and those produced by pointwise methods is not conducted. This is because our reviews are designed for breadth-oriented, parallel assessment of contemporaneously published literature, whereas pointwise methods primarily perform deep, fine-grained analysis of a single work. These methodological differences result in a fundamental mismatch in evaluation scope and objectives, making direct comparison inappropriate. We encourage combining our breadth-oriented reviews with pointwise evaluations to produce review content that is more comprehensive, better optimized, and more robust.

\begin{table}[t]
	\centering
	\scriptsize
	\setlength{\tabcolsep}{3pt}
	\begin{tabularx}{\columnwidth}{@{}
			>{\raggedright\arraybackslash}p{0.50\linewidth}
			>{\raggedright\arraybackslash}p{0.25\linewidth}
			>{\raggedright\arraybackslash}p{0.25\linewidth}
			@{}}
		\toprule
		& \textbf{DeepReview} & \textbf{CNPE} \\
		\midrule
		Computational complexity (dense) & $\displaystyle O(n L^2)$ & $\displaystyle O(\alpha n^2 L^2)$ \\
		$n$ & 634 & 634 \\
		$\alpha$ & — & 0.05 \\
		$L_{\mathrm{input}}$ & 10714.2 & 772.7 \\
		$L_{\mathrm{output}}$ & 12259.6 & 339.2 \\
		$L$ & 22973.8 & 1111.9 \\
		\midrule
		\textbf{Token cost} & 1$\times$ & 1.53$\times$ \\
		\textbf{Computational cost} & 1$\times$ & 0.074$\times$ \\
		\bottomrule
	\end{tabularx}
	\caption{Comparison between DeepReview and our method on token cost and computational cost.}
	\label{tab:deepreview_vs_ours}
\end{table}

\begin{table*}[t]
	\centering
	\scriptsize
	\setlength{\tabcolsep}{3pt}
	\begin{tabularx}{\textwidth}{@{}
			>{\raggedright\arraybackslash}p{0.16\linewidth}
			*{9}{>{\centering\arraybackslash}X}@{}} 
		\toprule
		& \multicolumn{4}{c}{\textbf{Decision}} & \multicolumn{4}{c}{\textbf{Ranking}} &  \\
		\cmidrule(lr){2-5} \cmidrule(lr){6-9}
		\textbf{Method} & \textbf{Accuracy} & \textbf{F1} & \textbf{AUC} & \textbf{Cohen \(\kappa\)} & \textbf{Spearman \(\rho\)} & \textbf{Pair. Acc.} & \textbf{MAP@20} & \textbf{NDCG@20} & \textbf{Avg. Perf.} \\
		\midrule
		\multicolumn{10}{@{}l}{\textbf{GLM}} \\
		Full Content (GLM) 
		& \textbf{0.6246} & \textbf{0.5630} & \textbf{0.6325} & \textbf{0.1261} 
		& \textbf{0.3018} & \textbf{0.6066} & \underline{0.3474} & \underline{0.7396} & \textbf{0.9446} \\
		Title \& Abs.\ (GLM) 
		& \underline{0.6183} & \underline{0.5557} & \underline{0.5893} & \underline{0.1114} 
		& \underline{0.2052} & \underline{0.5723} & \textbf{0.5576} & \textbf{0.7919} & \underline{0.9269} \\
		\midrule
		\multicolumn{10}{@{}l}{\textbf{Gemini}} \\
		Full Content (Gemini) 
		& \textbf{0.6246} & \textbf{0.5630} & \textbf{0.6054} & \textbf{0.1261} 
		& \textbf{0.2353} & \textbf{0.5837} & \underline{0.2920} & \textbf{0.7522} & \underline{0.9620} \\
		Title \& Abs.\ (Gemini) 
		& \underline{0.6215} & \underline{0.5594} & \underline{0.5975} & \underline{0.1187} 
		& \underline{0.2259} & \underline{0.5789} & \textbf{0.4196} & \underline{0.7517} & \textbf{0.9835} \\
		\bottomrule
	\end{tabularx}
	\caption{Comparison between Full Content and Titles + Abstracts. }
	\label{tab:glm_gemini_comparison}
\end{table*}

\section{Efficiency Analysis}
\label{sec:efficiency_analysis}

Building on our experimental results, we conducted a comparative analysis between DeepReview and our proposed approach. The outcomes are summarized in Table \ref{tab:deepreview_vs_ours}, focusing on total token consumption and estimated computational cost. 

For both methods, we randomly sampled 200 papers and computed the mean token usage to estimate per‑paper consumption. Computational cost was assessed under the assumption of dense architectures, as both our model (based on Qwen2.5-7B) and DeepReview (based on Phi-4 14B) adopt dense designs. We disregarded parameter‑scale differences, and given the use of global attention mechanisms in both models, the cost scales quadratically with context length.

Based on the computational complexity analysis, our approach consumes slightly more tokens than DeepReview (1.5×) yet achieves a dramatically lower actual computational cost (0.074×) with dense architectures. This reduction stems from the short‑inference scheme employed in our model, which not only yields superior performance but also substantially decreases computation overhead. Additionally, the shorter context length reduces memory usage, enabling deployment on consumer‑grade GPUs. These results highlight that efficiency gains can be achieved without sacrificing task quality.

\section{Justification for Evaluation Scope}

Limiting evaluation to titles and abstracts is a justified trade-off, informed by both prior literature and the characteristics of comparison-based systems. First, full content has been shown to provide limited information for assessing novelty and core contributions \cite{zhou2024llm}. Second, information beyond titles and abstracts has only marginal impact on the performance of comparison-based paper evaluation systems \cite{hopner2025automatic}. Finally, using only titles and abstracts for comparison-based systems is a widely accepted practice \cite{zhao2025words}.

Comparison-based systems differ from pointwise approaches in their goals and data usage. Our focus on titles and abstracts reflects how scholars perceive contemporaneous developments. The system emphasizes breadth and domain relationships rather than deep technical details. Our intention is to provide complementary insights for reviewers, not to replace pointwise systems that rely on full text. Instead, we aim to establish comparison-based approaches as an effective supplement.

To evaluate the sufficiency of titles and abstracts, we implemented a baseline comparison strategy \cite{zhang2025from} without augmentation or training, testing both full content and titles + abstracts (see Table \ref*{tab:glm_gemini_comparison}):  
(1) GLM results: Paired $t$-test $p = 0.434$. Full content yields a slight average improvement, but differences are not statistically significant; 
(2) Gemini results: Paired $t$-test $p = 0.792$. Titles + abstracts perform slightly better on average, with still no significant difference.
These results show that, in comparison-based evaluation, access to the full text offers limited benefit compared with improvements from data augmentation or model optimization. 

This setup also provides several advantages: (1) Robustness under partial information: Strong performance without full text demonstrates the adequacy and resilience of our design; 
(2) Computational efficiency: Processing only titles and abstracts reduces resource demands;
(3) Real-world applicability: In domains where full text is often unavailable, this approach is more deployable.

Based on both literature evidence and experimental data, we argue that using titles and abstracts alone is a reasonable and effective choice. Nevertheless, we acknowledge that this inevitably excludes certain methodological and experimental details, and we warn readers of the associated risks and bias. We hope that the additional information provided will help readers comprehensively view the advantages and disadvantages of our method and take them into consideration while using it.

\section{Visualization}
\label{sec:visualization}
\begin{figure}[htbp]
	\centering
	\includegraphics[width=\columnwidth]{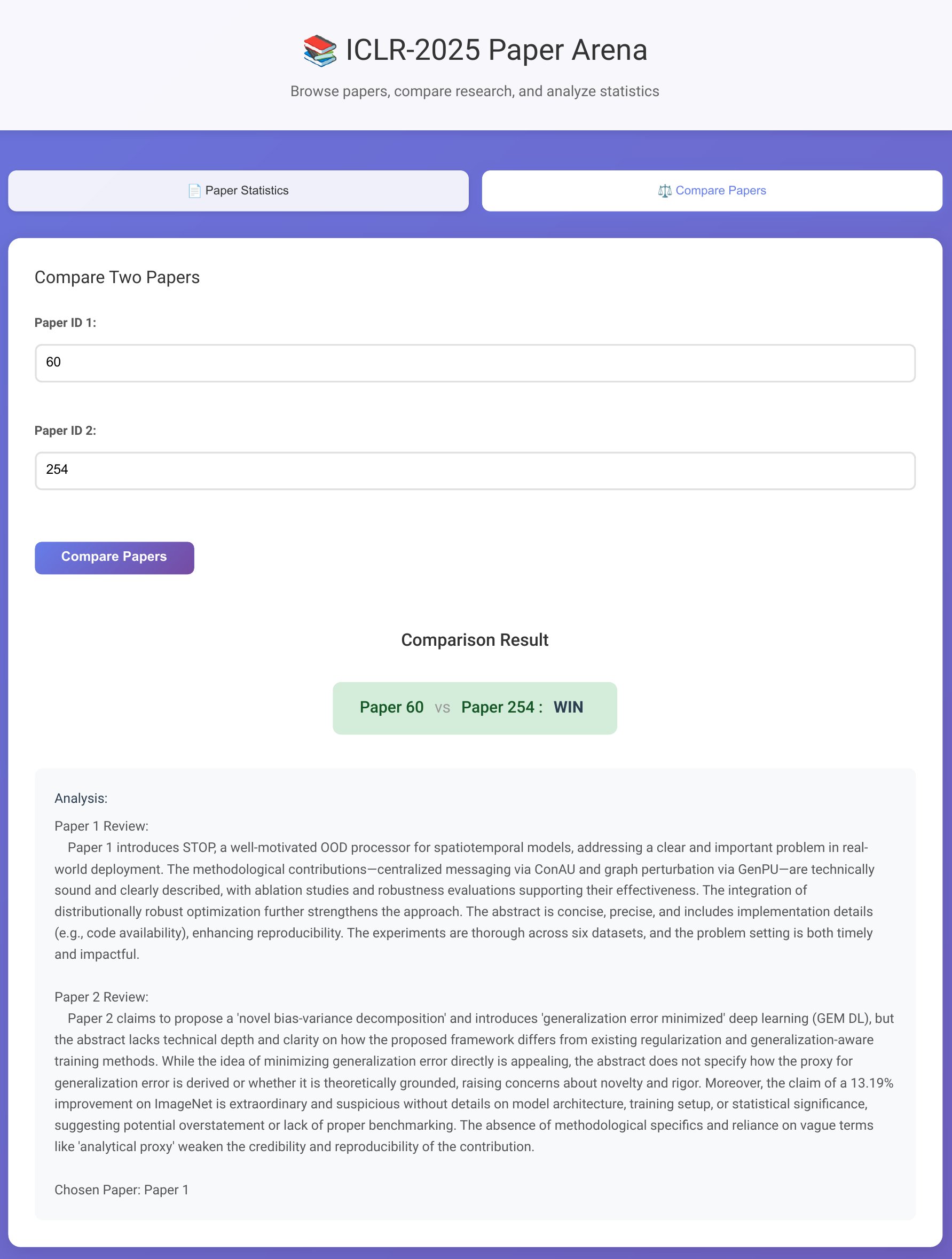}
	\caption{Paper Comparison Analysis View}
	\label{fig:web_double}
\end{figure}

\begin{figure}[htbp]
	\centering
	\includegraphics[width=\columnwidth]{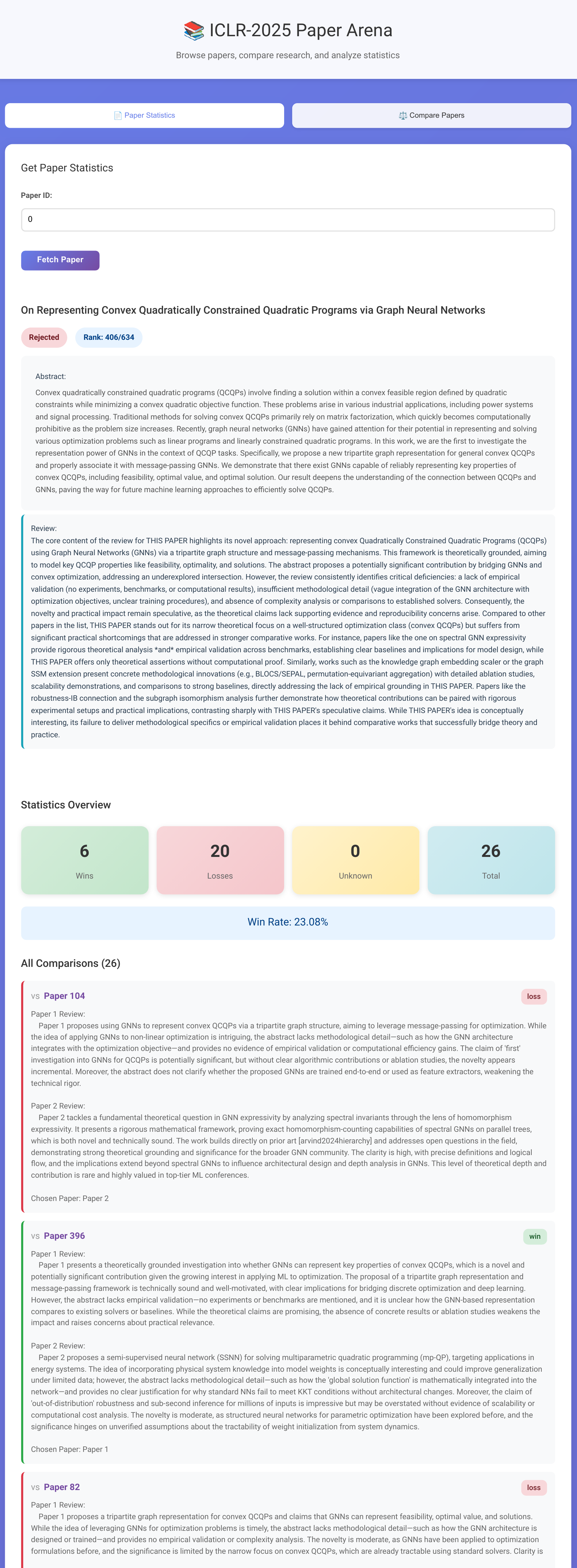}
	\caption{Single Paper Statistic View}
	\label{fig:web_single}
\end{figure}

To facilitate an intuitive understanding of our method's performance on the ICLR-2025 test set and to enable fine-grained comparisons among different submissions, we built an interactive demonstration system. The front end is implemented in Vue.js, and the back end is powered by FastAPI. The system offers two main functionalities:
(1) Single Paper Statistic View (See Figure \ref{fig:web_single}). Given a paper ID, the system presents a comprehensive analysis of the paper, including its metadata, simulated review, overall ranking within the test set, prediction outcomes, comparative statistics against other papers, and aggregated win rate.
(2) Paper Comparison Analysis View (See Figure \ref{fig:web_double}). Given two paper IDs, the model independently reviews each paper and provides a comparative assessment of their relative merits, along with explanatory text.
This demo is intended to serve as an interactive and interpretable evaluation platform, enabling researchers to better understand the model's behavior and outcomes in realistic conference review scenarios.

\end{document}